\documentclass[sigplan,screen,letterpaper]{acmart}

\usepackage{array}
\usepackage{float}
\usepackage{hyperref}
\usepackage{listings}
\usepackage{xcolor}

\newcommand{\rev}[1]{\textcolor{black}{#1}}

\usepackage{tikz}
\usetikzlibrary{
  arrows.meta,
  positioning,
  fit,
  backgrounds,
  calc,
  shapes.geometric,
  decorations.pathreplacing
}

\definecolor{srccolor}{HTML}{E8F0FE}
\definecolor{mlircolor}{HTML}{D5F5E3}
\definecolor{ourcolor}{HTML}{FDEBD0}
\definecolor{ourborder}{HTML}{E67E22}
\definecolor{llvmcolor}{HTML}{E8DAEF}
\definecolor{dslcolor}{HTML}{FCF3CF}
\definecolor{dslborder}{HTML}{F1C40F}
\definecolor{defaultcolor}{HTML}{D6DBDF}
\definecolor{defaultborder}{HTML}{808B96}
\definecolor{rtcolor}{HTML}{D5F5E3}
\definecolor{rtborder}{HTML}{27AE60}

\AtBeginDocument{%
  }

\setcopyright{none} 
\copyrightyear{2027}
\acmYear{2027}
\acmDOI{XXXXXXX.XXXXXXX}

\acmConference[CGO 2027]{The IEEE/ACM International Symposium on Code Generation and Optimization}{March 2027}{Salt Lake City, UT, USA}

\acmISBN{978-X-XXXX-XXXX-X/XX/XX}

\begin{document}

\title{OpenMP Meta-Lowering: A Declarative Approach to Performance Portable Parallel Code Generation}


\author{Luca Parigi}
\affiliation{%
  \institution{University of Bologna}
  \city{Bologna}
  \state{BO}
  \country{Italy}
}
\email{luca.parigi2@unibo.it}

\author{Giuseppe Tagliavini}
\affiliation{%
  \institution{University of Bologna}
  \city{Bologna}
  \state{BO}
  \country{Italy}
}
\email{giuseppe.tagliavini@unibo.it}



    
\begin{abstract}

The increasing diversity of parallel hardware challenges existing compilation flows.
While OpenMP provides a portable abstraction for shared-memory parallelism, existing compilers tightly couple the frontend semantics with fixed lowering strategies. This design limits performance portability across different runtimes and architectures.
In this paper, we present a modular approach to parallel code generation based on the Multi-Level Intermediate Representation (MLIR) framework. 
Our approach combines an OpenMP frontend with a domain-specific language (DSL) that specifies how OpenMP constructs are lowered for a target. We call this approach OpenMP meta-lowering, treating lowering as a programmable component rather than compiler-specific logic.
This design provides explicit control over code outlining, data sharing, and runtime interfacing across diverse targets.
We evaluate our approach on PolyBench/C-OMP across general-purpose and embedded multicore targets.
Our results show that the proposed method matches the performance of state-of-the-art compiler toolchains with negligible code-size overhead (< 0.7\%). 
\rev{Preserving OpenMP constructs until late lowering stages enables both general-purpose and OpenMP-specific MLIR optimizations.}
By decoupling lowering from compiler internals, our approach \rev{expresses the same OpenMP subset in \rev{5,182} lines of code, $\sim$32\% fewer than the corresponding lowering code in Clang and  $\sim$76\% fewer than in GCC, while supporting the pmsis runtime costs only a few specification lines, enabling rapid support for new runtimes without modifying the compiler}.


\end{abstract}



\maketitle

\section{Introduction}
The increasing heterogeneity of modern computing platforms, ranging from general-purpose CPUs to embedded systems, has made performance portability a key requirement, as maintaining separate implementations for each target architecture significantly increases development time and cost~\cite{deakin2019perfportability}.
This challenge is particularly critical in parallel computing systems, where differences in execution models, memory hierarchies, and synchronization mechanisms complicate the development of software that enforces portability without sacrificing efficiency~\cite{diaz2012survey}.

To address this complexity, high-level parallel programming models have been introduced to allow developers to express parallelism independently of the underlying hardware. Open Multi-Processing (OpenMP) has emerged as one of the most widely used models for shared-memory parallel platforms~\cite{dagum1998openmp,openmp_spec}. It provides a directive-based abstraction that enables developers to express parallelism and data-sharing semantics through source-code annotations.
By abstracting low-level aspects of parallel execution, such as thread creation, synchronization, and workload distribution, OpenMP allows developers to write portable parallel code through a set of compiler directives and runtime services. This approach has been demonstrated to be effective across various domains, including high-performance computing~\cite{kadosh2023openmp}, desktop applications, and embedded systems~\cite{rossi2014energy}.

The portability of OpenMP is mainly limited to the source code level. The translation of OpenMP constructs into executable code relies on compiler-specific strategies that are closely tied to a particular runtime implementation. Modern toolchains, such as LLVM and GCC, convert OpenMP directives into calls to their respective runtime libraries (e.g., \emph{libomp} for LLVM and \emph{libgomp} for GCC). Consequently, assumptions about runtime interfaces and low-level execution semantics are embedded directly into the compiler infrastructure, limiting portability across alternative runtimes and execution environments.
This tight coupling is evident even for simple constructs. For instance, consider a \texttt{\#pragma omp parallel for} loop: depending on the target runtime, the compiler must generate different sequences of runtime calls, such as \texttt{\_\_kmpc\_fork\_call} and \texttt{\_\_kmpc\_for\_static\_init} for \emph{libomp}~\cite{libomp}, or \texttt{GOMP\_parallel} followed by explicit loop-boundary computations for \emph{libgomp}~\cite{libgomp}, respectively.
While these implementations achieve the same high-level semantics, their interfaces and calling conventions differ significantly. In current implementations, the compiler integrates runtime-specific logic directly into the lowering pipeline, making it difficult and error-prone to retarget to new runtimes.

This approach becomes particularly challenging when targeting embedded and ultra-low-power architectures. While general-purpose platforms rely on mature runtimes, embedded systems often require specialized runtime implementations that sacrifice generality for efficiency~\cite{chapman2009openmp,wang2013libeomp,rossi2014energy}. Supporting these environments typically requires re-implementing significant parts of the lowering pipeline to align with the target application binary interface (ABI) and execution model. Consequently, expanding OpenMP support to new architectures or runtimes demands considerable engineering effort and restricts the ability to experiment with different execution strategies~\cite{AGATHOS2022102895}.

To address the aforementioned limitations, this paper presents a modular approach to OpenMP lowering across multiple targets.
The proposed methodology introduces a meta-layer on top of the Multi-Level Intermediate Representation (MLIR)~\cite{lattner2020mlir,lattner2021mlir}, a compiler infrastructure that enables flexible program representations at different levels of abstraction.
However, despite this flexibility, the current approach to lowering OpenMP constructs within the MLIR infrastructure remains tightly coupled to specific runtime implementations.
The key insight of this work is that the lowering process should not be hardwired into the compiler, but instead expressed as a declarative mapping between high-level parallel semantics and runtime interfaces.
We reformulate the lowering process as a programmable component by introducing a domain-specific language (DSL) that describes lowering rules for diverse runtime implementations as an external specification, decoupling parallel semantics from runtime-specific details.
Unlike previous MLIR-based approaches, which rely on fixed lowering pipelines~\cite{lopoukhine2025snitch} or imperative transformations implemented through compiler passes and pattern rewrites~\cite{mlir_drr, mlir_pdll}, our methodology treats lowering as a declarative, runtime-parametric abstraction.
To the best of our knowledge, this is the first approach to make OpenMP lowering explicitly programmable and runtime-agnostic within a production compiler infrastructure.
Our implementation is available as open-source software\footnote{\url{https://github.com/EEESlab/mlir-opt-omp}}, with an archived snapshot\footnote{\url{https://zenodo.org/records/22211483}}.

\sloppypar
Our approach replaces the traditional monolithic lowering pipeline with a three-phase process: annotating the Intermediate Representation (IR), outlining functional contexts, and applying the lowering plan.
The lowering process is driven by the meta-level specification expressed through DSL rules, enabling each phase to be customized for different runtime targets. This design fully decouples OpenMP semantics from any particular runtime implementation, allowing the same high-level representation to be lowered to multiple backends without embedding runtime-specific logic into the compiler infrastructure.

To summarize, this paper presents the following contributions:
\begin{itemize}
\item A meta-lowering methodology that expresses the OpenMP lowering by means of a DSL specification, interpreted by the compiler infrastructure in a runtime-agnostic and extensible manner.
\item A modular three-phase MLIR lowering pipeline that instantiates the runtime-specific behavior based on the rules provided by the DSL specification.
\item An end-to-end integration of the DSL and MLIR pipeline into a Clang/CIR-based compiler flow, enabling the combination with additional MLIR transformations from very early compilation stages.
\item An evaluation on the PolyBench/C-OMP benchmark suite, across general-purpose and embedded targets, showing that our approach matches the performance of state-of-the-art compiler toolchains while incurring negligible code size overhead. \rev{Additionally, the evaluation considers the effects of illustrative general-purpose and OpenMP-specific optimizations}. 
\end{itemize}

We demonstrate the flexibility of our approach by targeting heterogeneous execution environments, including general-purpose runtimes (\emph{libomp}, \emph{libgomp}) and a lightweight embedded runtime for the Parallel Ultra-Low Power (PULP) multicore platform~\cite{conti2025pulp}.
Across the evaluated benchmarks, our approach matches the performance of state-of-the-art compiler toolchains while introducing negligible code size overhead (below 0.7\%).
This decoupling \rev{reduces engineering complexity: the whole lowering pipeline takes 5,040 lines of code, $\sim$32\% less than the corresponding lowering code in Clang and $\sim$76\% less than in GCC for the same OpenMP subset, and adding support for a new runtime costs a few lines in the specification.}


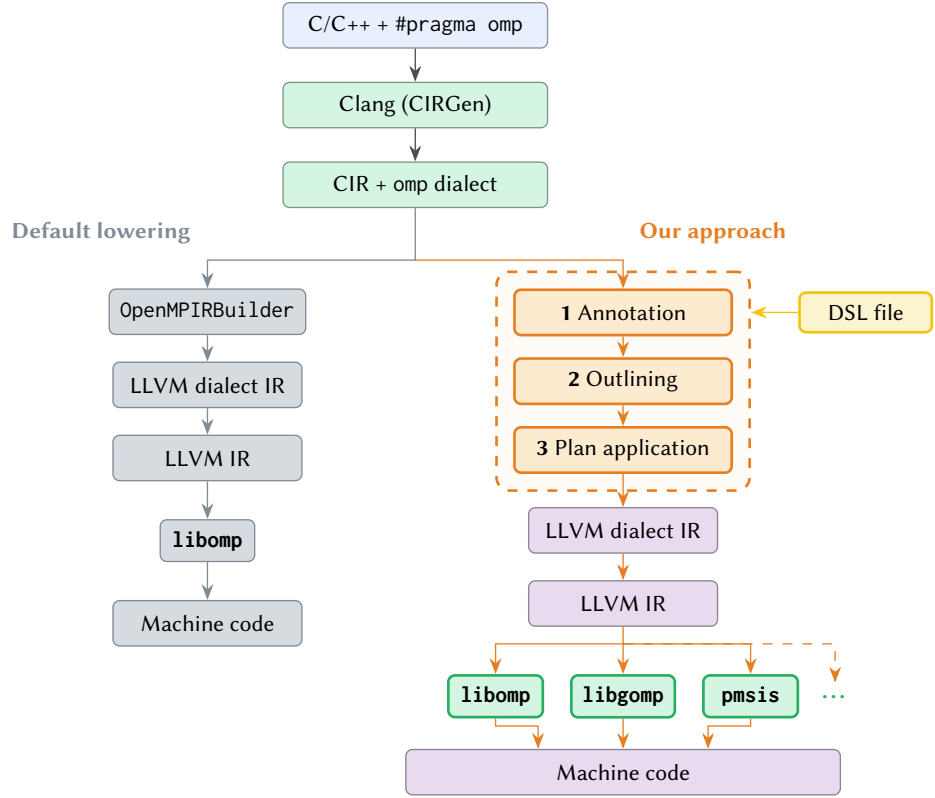
\begin{figure*}[t]
\centering
\resizebox{0.70\textwidth}{!}{%
\begin{tikzpicture}[
    node distance=0.3cm,
    box/.style={
        draw, rounded corners=2pt, minimum width=2.0cm,
        minimum height=0.48cm, align=center,
        font=\scriptsize\sffamily
    },
    phase/.style={
        box, fill=ourcolor, draw=ourborder, thick,
        minimum width=2.3cm
    },
    stdbox/.style={
        box, fill=mlircolor, draw=mlircolor!60!black
    },
    llvmbox/.style={
        box, fill=llvmcolor, draw=llvmcolor!60!black
    },
    srcbox/.style={
        box, fill=srccolor, draw=srccolor!60!black,
        minimum width=2.8cm
    },
    defbox/.style={
        box, fill=defaultcolor, draw=defaultborder
    },
    dslbox/.style={
        box, fill=dslcolor, draw=dslborder, thick,
        minimum width=1.4cm, minimum height=0.42cm,
        font=\scriptsize\sffamily
    },
    rtbox/.style={
        draw, rounded corners=2pt, minimum width=1.0cm,
        minimum height=0.4cm, align=center,
        font=\scriptsize\sffamily\bfseries,
        fill=rtcolor, draw=rtborder, thick
    },
    defrt/.style={
        draw, rounded corners=2pt, minimum width=1.0cm,
        minimum height=0.4cm, align=center,
        font=\scriptsize\sffamily\bfseries,
        fill=defaultcolor, draw=defaultborder
    },
    arr/.style={-Stealth, thin, color=black!70},
    defarr/.style={-Stealth, thin, color=defaultborder},
    ourarr/.style={-Stealth, thin, color=ourborder}
]

\node[srcbox] (src)
    {C/C++ + \texttt{\#pragma omp}};
\node[stdbox, below=0.35cm of src, minimum width=2.8cm] (clang)
    {Clang (CIRGen)};
\node[stdbox, below=0.35cm of clang, minimum width=2.8cm] (cirop)
    {CIR + \texttt{omp} dialect};

\draw[arr] (src)   -- (clang);
\draw[arr] (clang) -- (cirop);

\coordinate[below=0.55cm of cirop] (branch);

\coordinate (leftarm)  at ($(branch) + (-2.2, 0)$);
\coordinate (rightarm) at ($(branch) + ( 2.2, 0)$);

\node[font=\scriptsize\sffamily\bfseries, text=defaultborder,
      anchor=south east,
      at={($(leftarm) + (-0.05, 0.05)$)}] (deflbl)
    {Default lowering};

\node[defbox, below=0.30cm of leftarm] (ompbuilder)
    {\texttt{OpenMPIRBuilder}};
\node[defbox, below=0.28cm of ompbuilder] (defllvm)
    {LLVM dialect IR};
\node[defbox, below=0.28cm of defllvm] (defir)
    {LLVM IR};
\node[defrt, below=0.4cm of defir] (libomp_def)
    {\texttt{libomp}};
\node[defbox, below=0.4cm of libomp_def] (defmcode)
    {Machine code};

\draw[defarr] (cirop.south) -- (branch) -- (leftarm)
    -- (ompbuilder.north);
\draw[defarr] (ompbuilder) -- (defllvm);
\draw[defarr] (defllvm)    -- (defir);
\draw[defarr] (defir)      -- (libomp_def);
\draw[defarr] (libomp_def) -- (defmcode);

\node[font=\scriptsize\sffamily\bfseries, text=ourborder,
      anchor=south west,
      at={($(rightarm) + (0.05, 0.05)$)}] (ourlbl)
    {Our approach};

\node[phase, below=0.30cm of rightarm] (p1)
    {\textbf{1} Annotation};
\node[phase, below=0.22cm of p1] (p2)
    {\textbf{2} Outlining};
\node[phase, below=0.22cm of p2] (p3)
    {\textbf{3} Plan application};
\node[llvmbox, below=0.35cm of p3] (ourllvm)
    {LLVM dialect IR};
\node[llvmbox, below=0.28cm of ourllvm] (ourir)
    {LLVM IR};

\draw[ourarr] (branch) -- (rightarm) -- (p1.north);
\draw[ourarr] (p1) -- (p2);
\draw[ourarr] (p2) -- (p3);
\draw[ourarr] (p3) -- (ourllvm);
\draw[ourarr] (ourllvm) -- (ourir);

\begin{scope}[on background layer]
\node[
    draw=ourborder, thick, dashed, rounded corners=4pt,
    fill=ourcolor!20,
    fit=(p1)(p2)(p3),
    inner xsep=5pt, inner ysep=5pt,
] (ourfit) {};
\end{scope}

\node[dslbox, right=0.5cm of ourfit.east |- p1] (dsl)
    {DSL file};
\draw[arr, dslborder] (dsl.west) -- (ourfit.east |- p1);

\node[rtbox, below=0.5cm of ourir, xshift=-1.35cm] (rt1)
    {\texttt{libomp}};
\node[rtbox, below=0.5cm of ourir] (rt2)
    {\texttt{libgomp}};
\node[rtbox, below=0.5cm of ourir, xshift=1.35cm] (rt3)
    {\texttt{pmsis}};
\node[right=0.12cm of rt3,
      font=\scriptsize\bfseries, text=rtborder] (dots)
    {\ldots};

\coordinate[below=0.18cm of ourir] (fanout);
\draw[ourarr] (ourir.south) -- (fanout) -| (rt1.north);
\draw[ourarr] (fanout) -| (rt2.north);
\draw[ourarr] (fanout) -| (rt3.north);
\draw[ourarr, dashed] (fanout) -| (dots.north);

\node[llvmbox, below=0.32cm of rt2, minimum width=4.6cm] (ourmcode)
    {Machine code};
\draw[ourarr] (rt1.south) -- ++(0,-0.08)
    -| ([xshift=-0.9cm]ourmcode.north);
\draw[ourarr] (rt2.south) -- (ourmcode.north);
\draw[ourarr] (rt3.south) -- ++(0,-0.08)
    -| ([xshift=0.9cm]ourmcode.north);

\end{tikzpicture}
}

\caption{Compilation pipeline overview. The default LLVM/Clang flow targets only libomp, whereas our three-phase DSL-driven pipeline supports multiple runtimes and can be extended to additional backends.}
\label{fig:pipeline}
\Description{Diagram comparing two compilation pipelines. The left side shows the default LLVM/Clang flow, where OpenMP constructs are lowered directly to LLVM IR targeting libomp. The right side shows the proposed approach with three phases—annotation, outlining, and plan application—guided by a DSL file, enabling lowering to multiple runtimes such as libomp, libgomp, and PMSIS.}
\end{figure*}

\section{Background}

Our work builds on MLIR and the OpenMP compilation infrastructure. MLIR provides a multi-level intermediate representation in which high-level program semantics can be preserved and progressively lowered toward machine code. In particular, OpenMP constructs are represented through the dedicated \texttt{omp} dialect and coexist with other dialects, such as ClangIR (CIR), within the same compilation pipeline.

Figure~\ref{fig:pipeline} contrasts the conventional OpenMP compilation flow with our approach. In the standard LLVM/Clang pipeline, OpenMP operations are eventually translated through OpenMPIRBuilder into calls to the LLVM OpenMP runtime (\texttt{libomp}). As a result, runtime-specific assumptions are embedded in the lowering implementation. In contrast, our approach inserts a configurable meta-lowering layer that drives lowering via an external DSL specification, enabling the same OpenMP representation to target multiple runtimes without modifying compiler internals.

\subsection{MLIR-Based OpenMP Compilation}

MLIR~\cite{lattner2020mlir} is a compiler infrastructure designed to support multiple levels of abstraction within a single compilation framework. Rather than lowering source code directly to a low-level representation, MLIR organizes programs into \emph{dialects}, each providing domain-specific operations, types, and transformations. This design enables high-level program information to be preserved throughout compilation and progressively lowered toward machine code.

Our work builds on ClangIR (CIR)~\cite{clangir}, a high-level MLIR dialect for C/C++. CIR represents source-language constructs such as functions, loops, variables, and control-flow structures while retaining information that is typically lost during early lowering to LLVM IR. The CIR compilation flow integrates with the broader MLIR ecosystem, allowing CIR operations to coexist with operations from other dialects.

OpenMP support in MLIR is provided through the dedicated \texttt{omp} dialect, which models OpenMP directives and clauses. Since both CIR and OpenMP are represented as MLIR dialects, parallel constructs remain explicit alongside the surrounding C/C++ code throughout most of the compilation process. This separation between high-level parallel semantics and low-level code generation enables transformations and optimizations to be applied before runtime-specific lowering decisions are introduced.

\subsection{OpenMP Lowering}

OpenMP compilation requires translating high-level parallel constructs into runtime-specific implementations. This process typically involves three categories of transformations. First, \emph{function outlining} extracts the body of a parallel region into a separate function. Second, \emph{control-flow restructuring} rewrites work-sharing constructs, such as parallel loops, into forms compatible with the scheduling mechanisms provided by the target runtime. Third, \emph{runtime call emission} introduces calls that implement thread creation, synchronization, and work distribution.

In current MLIR-based toolchains, these transformations are implemented through OpenMPIRBuilder and are tightly coupled to the ABI of a specific runtime, namely \texttt{libomp}. For example, OpenMP constructs are lowered into runtime-specific calls such as \texttt{\_\_kmpc\_fork\_call} and loop scheduling routines whose interfaces are embedded directly into the lowering implementation. Consequently, supporting alternative runtimes requires modifying compiler-internal code rather than simply changing the runtime description.

This tight coupling between OpenMP semantics and runtime-specific lowering motivates the approach proposed in this paper. Instead of hardwiring runtime behavior into compiler passes, we externalize lowering decisions into a declarative specification that the compiler infrastructure can interpret and adapt to different runtime environments. 
\rev{A DSL specification defines which runtime calls to emit and which code-generation strategy to apply for each construct; the proposed compilation pipeline applies the actual changes.}


\section{Methodology}
In this section, we present our OpenMP meta-lowering approach. 
\rev{A DSL specification declares, for each OpenMP construct, how it is realized on a target runtime: the runtime calls to emit, their position relative to the construct body, and the handling of the body itself.
Section~\ref{sec:expressibility} describes what a specification can express and Section~\ref{sec:dsl-grammar} formalizes its grammar;
Section~\ref{sec:pipeline} describes the three-phase MLIR pipeline that performs the lowering, and Section~\ref{sec:dsl-semantics} walks through a complete example. The last two subsections state how correctness is checked and which constructs are currently supported.}

\subsection[What a specification can express]{\rev{What a specification can express}}
\label{sec:expressibility}
\rev{A specification outlines how each \texttt{omp}-dialect operation is
lowered for a target runtime. It does so in terms of four concepts:}

\begin{enumerate}
    \item \rev{The \texttt{omp}-dialect \emph{operations} that a specification can lower (\texttt{omp.parallel}, \texttt{omp.wsloop}, etc.).}
    \item \rev{The \emph{context variables} that each operation exposes, represented by the values of its clauses (\texttt{num\_threads},
    \texttt{nowait}).}
    \item \rev{The \emph{body-handling modes} available to a construct:
    its body may be outlined into a separate function, restructured as a
    work-sharing loop, or absent.}
    \item \rev{The \emph{structured rewrites} that a specification can invoke, each generating a dedicated code pattern.}
\end{enumerate}

\rev{There are no restrictions on the names of functions, which means that a specification can express any construct that reduces to a sequence of runtime calls associated with one of the body-handling modes. Section~\ref{sec:scope} outlines which OpenMP constructs are currently included in this set.}

\subsection{Lowering description language}
\label{sec:dsl-grammar}
The lowering process for a generic runtime is described using a DSL whose grammar is shown in Figure~\ref{fig:grammar}. A DSL program consists of one or more \emph{runtime declarations}, each of which bundles a set of lowering rules under a named runtime identifier. This design allows users to specify multiple runtimes in a single file and choose one at compile time.

\begin{figure}[t]
    \newcommand{\NT}[2]{\textsf{#1-#2}}
    \newcommand{\KW}[1]{\texttt{\textbf{#1}}}
    \newcommand{\TK}[1]{\textsf{\MakeUppercase{#1}}}

  \footnotesize
  \centering
  \noindent\rule{\linewidth}{0.5pt}

  \newlength{\grLHS}\setlength{\grLHS}{72pt}
  \newlength{\grREL}\setlength{\grREL}{22pt}
  \newlength{\grRHS}\setlength{\grRHS}{\linewidth}
  \addtolength{\grRHS}{-\grLHS}
  \addtolength{\grRHS}{-\grREL}
  \addtolength{\grRHS}{-9pt}

  \setlength{\tabcolsep}{3pt}
  \renewcommand{\arraystretch}{0.95}

  \begin{tabular}{%
    >{\sffamily}p{\grLHS}%
    >{$}c<{$}%
    >{\ttfamily}p{\grRHS}}
    program              & ::= & \{ \NT{runtime}{decl} \} \\[1pt]
    \NT{runtime}{decl}   & ::= & \KW{runtime}\ \TK{ident} \\
                         &     & \hspace{4pt}\{\ \{\ \NT{runtime}{item}\ \}\ \} \\[1pt]
    \NT{runtime}{item}   & ::= & \NT{let}{decl} \\
                         &     & \hspace{4pt}$|$\ \NT{construct}{decl} \\[1pt]
    \NT{construct}{decl} & ::= & \KW{construct}\ \TK{ident} \\
                         &     & \hspace{4pt}[\ \KW{when}\ \textit{predicate}\ ] \\
                         &     & \hspace{4pt}\{\ \{\ \NT{construct}{item}\ \}\ \} \\[1pt]
    \NT{construct}{item} & ::= & \NT{let}{decl} \\
                         &     & \hspace{4pt}$|$\ \NT{property}{decl} \\
                         &     & \hspace{4pt}$|$\ \NT{block}{decl} \\[1pt]
    \NT{property}{decl}  & ::= & \TK{ident}\ =\ \textit{expr}\ ; \\[1pt]
    \NT{block}{decl}     & ::= & (\ \KW{pre}\ $|$\ \KW{invoke}\ $|$\ \KW{post}\ ) \\
                         &     & \hspace{4pt}\{\ \{\ \textit{statement}\ \}\ \} \\[1pt]
    \NT{let}{decl}       & ::= & \KW{let}\ \TK{ident}\ =\ \textit{expr}\ ; \\[1pt]
    statement            & ::= & \KW{when}\ \textit{predicate}\ \texttt{=>}\ \textit{action}\ ; \\
                         &     & \hspace{4pt}$|$\ \KW{otherwise}\ \texttt{=>}\ \textit{action}\ ; \\
                         &     & \hspace{4pt}$|$\ \KW{let}\ \TK{ident}\ =\ \textit{expr}\ ; \\
                         &     & \hspace{4pt}$|$\ \textit{action}\ ; \\[1pt]
    action               & ::= & \KW{call}\ (\ \TK{ident}\ $|$\ \TK{string}\ ) \\
                         &     & \hspace{12pt}(\ [\ \NT{arg}{list}\ ]\ ) \\
                         &     & \hspace{4pt}$|$\ \KW{emit}\ \TK{ident} \\[1pt]
    predicate            & ::= & \NT{pred}{and}\ \{\ \KW{or}\ \NT{pred}{and}\ \} \\[1pt]
    \NT{pred}{and}       & ::= & \NT{pred}{not}\ \{\ \KW{and}\ \NT{pred}{not}\ \} \\[1pt]
    \NT{pred}{not}       & ::= & \KW{not}\ \NT{pred}{not}\ $|$\ \NT{pred}{atom} \\[1pt]
    \NT{pred}{atom}      & ::= & \KW{has}\ (\ \TK{ident}\ ) \\
                         &     & \hspace{4pt}$|$\ \TK{ident}\ \texttt{==}\ \textit{expr} \\
                         &     & \hspace{4pt}$|$\ \TK{ident}\ \texttt{!=}\ \textit{expr} \\
                         &     & \hspace{4pt}$|$\ \TK{ident}\ $|$\ \KW{true}\ $|$\ \KW{false} \\
                         &     & \hspace{4pt}$|$\ (\ \textit{predicate}\ ) \\[1pt]
    expr                 & ::= & \TK{ident}\ (\ [\ \NT{arg}{list}\ ]\ ) \\
                         &     & \hspace{4pt}$|$\ \TK{ident}\ $|$\ \TK{string}\ $|$\ \TK{number} \\
                         &     & \hspace{4pt}$|$\ \KW{true}\ $|$\ \KW{false} \\
                         &     & \hspace{4pt}$|$\ [\ [\ \NT{arg}{list}\ ]\ ]\ $|$\ (\ \textit{expr}\ ) \\[1pt]
    \NT{arg}{list}       & ::= & \textit{expr}\ \{\ ,\ \textit{expr}\ \} \\
  \end{tabular}

  \vspace{2pt}
  \noindent\rule{\linewidth}{0.5pt}

  \caption{EBNF grammar of the DSL.}
  \label{fig:grammar}
  \Description{EBNF grammar table of the lowering DSL. The grammar defines 20 non-terminals organized into three groups. The top-level structure consists of a program made of runtime declarations, each containing let declarations and construct declarations. A construct declaration maps a named OpenMP IR construct to lowering rules and optionally includes a predicate guard. Its body is a list of property declarations, let declarations, and block declarations. Block declarations are organized into three lifecycle hooks: pre, invoke, and post, each containing a sequence of statements. A statement is either a conditional action guarded by a when or otherwise clause, a let binding, or an unconditional action. Actions are either a call, which invokes a named or dynamically resolved function with an argument list, or an emit, which triggers a named IR transformation pass. Predicates support conjunction, disjunction, and negation with standard precedence, and atomic predicates test attribute presence, equality, or inequality. Expressions cover function calls, identifiers, string, integer, and boolean literals, lists, and parenthesized sub-expressions. Primitive token types IDENT, STRING, and NUMBER are supplied by the lexer and are not expanded in the grammar.}
\end{figure}

\paragraph{Runtime and construct declarations.}
A \texttt{runtime} block may contain local bindings (\emph{let-decl}) and any number of \emph{construct declarations}. Each construct declaration maps an \rev{operation of the \texttt{omp} dialect} to a set of lowering rules. 
An optional \texttt{when} guard, expressed as a predicate over the construct attributes, allows the same construct to be handled differently depending on runtime conditions. For example, it can be used to differentiate between parallel regions that specify or do not specify a \texttt{num\_threads} clause.

\paragraph{Construct items.}
A construct declaration consists of a flat list that includes three types of items. A \emph{property declaration} associates a \rev{name} with an expression, making it available as a symbol during the lowering process. A \emph{let declaration} introduces a constant binding that is valid within the scope of the construct. A \emph{block declaration} describes the lowering process in terms of three lifecycle 
\rev{blocks}: \texttt{pre}, executed before materializing the construct; \texttt{invoke}, which handles the materialization of the construct's parallel semantics and may either dispatch an outlined function via \texttt{call} or \rev{apply structured rewrites} via \texttt{emit}; and \texttt{post}, executed after the construct materialization.

\paragraph{Statements and actions.}
Within a lifecycle block, each statement either fires unconditionally or is guarded by a \texttt{when}/\texttt{otherwise} conditional. The executable unit of a statement is an \emph{action}, of which there are two forms. A \texttt{call} action invokes a function with a list of expressions as arguments, corresponding to the emission of a runtime library call into the lowered IR. 
\rev{An \texttt{emit} action names a structured rewrite supported by the infrastructure, such as \texttt{thread\_bounds} or \texttt{loop\_body}, allowing IR rewrites and runtime calls to be composed declaratively within the same lowering sequence.}


\paragraph{Predicates.}
Guards on both construct declarations and individual statements are expressed as \emph{predicates}. The grammar supports conjunction (\texttt{and}), disjunction (\texttt{or}), and negation (\texttt{not}), following standard precedence rules from lowest to highest. Atomic predicates can check for the presence of a named attribute using \texttt{has}, compare the attribute value with a reference using \texttt{==} or \texttt{!=}, or treat the attribute name as a boolean flag. Parentheses are available for explicit grouping.

\paragraph{Expressions.}
Expressions cover the value language used in property bindings, let declarations, and action arguments. An expression can be a function call, a bare identifier referring to a bound attribute or let variable, a string or integer literal, a boolean literal, or a list. Lists allow passing structured data, such as sets of clause names, as arguments to \texttt{call} or \texttt{emit} actions. Primitive token types (\textsf{IDENT}, \textsf{STRING}, \textsf{NUMBER}) are supplied by the lexer and are not expanded further in the grammar.

\subsection{MLIR Lowering Pipeline}
\label{sec:pipeline}
The lowering pipeline consists of three MLIR passes orchestrated by \texttt{mlir-opt-omp}, a custom driver built on top of the standard \texttt{MlirOptMain} entry point. The driver registers the dialects required throughout the pipeline: \texttt{cir}, \texttt{arith}, \texttt{func}, \texttt{llvm}, \texttt{omp}, and the custom \texttt{omp\_lower} dialect.
Two flags control the global lowering configuration: \texttt{--omp-lower-dsl} specifies the path to the DSL file containing the runtime declarations, and \texttt{--omp-lower-runtime} selects the target runtime from that file.
Dialect legalization is handled transparently by the MLIR conversion framework, which verifies that all operations are legal with respect to the registered dialects after each pass.
Figure~\ref{fig:pipeline} shows the three passes and their integration into the compilation pipeline.

\paragraph{Pass 1: Annotation (\texttt{--omp-to-omp-lower}).}
The first pass processes the DSL file using a dedicated parser and evaluates the lowering declarations for the selected runtime.
\rev{For each visited operation, the pass selects the declaration to apply based on the actual values of the clauses: a \texttt{when} guarded declaration takes precedence over an unguarded one. When no declaration matches, the pass reports an error.}
\rev{Based on the selected declaration, the pass then attaches} the lowering plan to \rev{the} \texttt{omp} dialect operation, encoded as an additional operation belonging to the custom \texttt{omp\_lower} dialect. This operation records the target runtime name, the construct identifier, a dictionary of resolved property bindings, and three data structures carrying the \texttt{pre}, \texttt{invoke}, and \texttt{post} action sequences.
As a result of this pass, the structure of the \texttt{omp} operations remains unchanged, but they now carry all the necessary meta-level information required by the subsequent passes.

\paragraph{Pass 2: Outlining (\texttt{--omp-outline}).}
\rev{The second pass considers the lowering annotations from Pass~1 and realizes the body-handling mode that each construct requires. Where that mode is function outlining (e.g., \texttt{omp.parallel}), it} collects the variables crossing the region boundary, extracts the body into a separate function, and rewrites the original site into a dispatch point.
The signature of the outlined function and the way captured variables are passed to it are not hardcoded in the pass, but dictated by \texttt{outline\_signature} and \texttt{capture\_strategy} properties of the plan.
\rev{When using \texttt{by\_pointer}, each capture is treated as a separate argument in the outlined function; with \texttt{packed}, captures are combined into a single structure allocated at the call site, which defines the \texttt{env\_size} and \texttt{env\_align} context variables for size and alignment, respectively.}
\rev{For an \texttt{omp.wsloop}, whose body is restructured, the pass computes the iteration range assigned to each thread and rebuilds the loop over that range.}

\paragraph{Pass 3: Plan application (\texttt{--omp-lower-plan}).}
The third pass applies the lowering plan produced in Pass~1 to the outlined code from Pass~2, materializing the \texttt{pre}, \texttt{invoke}, and \texttt{post} action sequences for each construct.
For each \texttt{omp} operation, the pass first processes the \texttt{pre} actions, inserting the corresponding \texttt{llvm} dialect calls or \rev{applying structured rewrites} immediately before the construct in the IR.
It then processes the \texttt{invoke} actions, which materialize the construct at the point where the original \texttt{omp} operation appears.
Finally, it processes the \texttt{post} actions, inserting any required teardown or synchronization code immediately after the construct.

For each set of actions, \texttt{call} actions are lowered to direct \texttt{llvm.call} operations with the resolved function name and argument list.
\rev{Each argument name resolves to an SSA value in the IR: \texttt{num\_threads} and \texttt{if\_clause} to operands that the \texttt{omp} operation already carries, \texttt{body} to the function outlined by Pass~2, and \texttt{env\_ptr} to the capture structure built at the dispatch site.}
\texttt{emit} actions are dispatched by name to a registry of standard IR rewrite patterns, each of which produces a fixed sequence of \texttt{llvm} dialect operations, detailed in Section~\ref{sec:dsl-semantics}.

\subsection{DSL Semantics}
\label{sec:dsl-semantics}

\rev{This section walks through a concrete DSL specification, then characterizes the \texttt{emit} primitives and the properties that parameterize them.}


\begin{figure}[t]

\lstdefinelanguage{LoweringDSL}{
keywords={runtime,construct,when,otherwise,let,call,emit,
         has,and,or,not,true,false,pre,invoke,post,if,goto},
sensitive=true,
morestring=[b]",
morecomment=[l]{//},
}
\lstdefinestyle{dslstyle}{
    language=LoweringDSL,
    basicstyle=\ttfamily\footnotesize,
    keywordstyle=\bfseries,
    stringstyle=\color{black},
    commentstyle=\itshape\color{gray},
    showstringspaces=false,
    columns=fixed,
    keepspaces=true,
    frame=tb,
    framerule=0.5pt,
    xleftmargin=4pt,
    xrightmargin=4pt,
    aboveskip=4pt,
    belowskip=4pt,
    numbers=left,
    numberstyle=\ttfamily\footnotesize\color{gray},
    numbersep=4pt,
}

\begin{lstlisting}[style=dslstyle]
runtime libgomp {
 construct parallel {
  outline_signature = closure(env_ptr);
  capture_strategy  = "packed";
  invoke {
   when has(num_threads) =>
    call "GOMP_parallel"(
      body, env_ptr, num_threads, 0);
   otherwise =>
    call "GOMP_parallel"(body, env_ptr, 0, 0);
  }
 }
 construct wsloop when schedule == static {
  thread_id_function  = "omp_get_thread_num";
  num_thread_function = "omp_get_num_threads";
  pre    { emit thread_bounds; }
  invoke {
    emit loop_body;
  }
  post {
   when not nowait => call "GOMP_barrier"();
  }
 }
 construct wsloop when schedule == dynamic {
  chunk_start_function = "GOMP_loop_dynamic_start";
  chunk_next_function  = "GOMP_loop_dynamic_next";  
  pre {
   emit chunked_loop;
  }
  invoke {
   emit loop_body;
  }
  post {
   when nowait => call "GOMP_loop_end_nowait"();
   otherwise => call "GOMP_loop_end"();
  }
}
 construct task {
  outline_signature = closure(env_ptr);
  capture_strategy  = "packed";
  invoke {
   when has(if_clause) =>
    call "GOMP_task"(body, env_ptr, null,
                     env_size, env_align,
                     if_clause, 0, null, 0);
   otherwise =>
    call "GOMP_task"(body, env_ptr, null,
                     env_size, env_align,
                     true, 0, null, 0);
  }
 }
 construct barrier {
  invoke { call "GOMP_barrier"(); }
 }
}
\end{lstlisting}
\caption{Specification of the libgomp runtime (subset).}
\label{fig:dsl-example}
\Description{A DSL listing for the libgomp runtime defining lowering rules for four constructs. The parallel construct uses property declarations to fix the outline signature and capture strategy, and its invoke block conditionally calls GOMP_parallel with or without a thread count depending on whether a num_threads clause is present. Two wsloop constructs are defined, guarded by schedule == default and schedule == static respectively. The default variant emits thread bounds in the pre hook and the loop body in invoke, issuing a barrier in post unless nowait is set. The static variant calls GOMP_loop_static_start in pre to partition iterations, emits the loop body in invoke, and calls GOMP_loop_end followed by an optional barrier in post. The barrier construct reduces to a single GOMP_barrier call in its invoke hook.}
\end{figure}

\paragraph{A concrete example.}
Figure~\ref{fig:dsl-example} shows an excerpt of a DSL specification targeting the GCC OpenMP runtime (\emph{libgomp}), covering \rev{four} representative constructs. 
The \texttt{parallel} declaration \rev{opens with \texttt{outline\_signature} and \texttt{capture\_strategy} (lines~3 and~4), set to the libgomp convention: a single environment pointer, with the captures packed into one struct.}
These properties are consumed by the outlining pass before the \texttt{invoke} block is processed. 
The \texttt{invoke} block handles the dispatch: if the construct carries a \texttt{num\_threads} clause, tested with the \texttt{has} predicate, the outlined function is dispatched via \texttt{GOMP\_parallel} with the thread count as the third argument (line~7); otherwise, the argument is set to zero, signaling \emph{libgomp} to use the default thread count (line~9).
 
Two separate \texttt{wsloop} declarations handle the static and dynamic scheduling policies, selected at lowering time by the \texttt{when} guard on the construct declaration. 
For the static policy (lines~13--23), the \texttt{pre} \rev{block} emits a \rev{structured rewrite} (\texttt{thread\_bounds}) that computes the iteration range assigned to
each thread, and the \texttt{invoke} \rev{block} emits the loop body over that range. 
The \texttt{post} \rev{block} conditionally issues a \texttt{GOMP\_barrier} call unless the \texttt{nowait} clause is present, preserving OpenMP semantics. 
For the dynamic policy (lines~24--37), two properties name the dispatching functions, and the \texttt{post} \rev{block} selects the final call depending on \texttt{nowait}.

Similar to \texttt{parallel}, the \texttt{task} declaration provides the information required for outlining and generates a call to \texttt{GOMP\_task}, parameterized by a condition derived from one of the task clauses.

Finally, the \texttt{barrier} declaration (lines~52--54) illustrates the simplest lowering pattern: a single runtime call with no outlining and no control flow restructuring. 
The entire construct body reduces to one \texttt{GOMP\_barrier} call in the \texttt{invoke} \rev{block}, with \texttt{pre} and \texttt{post} absent.

\paragraph{Semantics.}
The rest of the semantics lies in the \texttt{emit} primitives and the properties that parameterize them, which fall into three classes:

\begin{enumerate}
\item \textbf{Runtime-agnostic parameters.} 
These properties parameterize a primitive describing a generic code-rewriting pattern.
The \texttt{thread\_bounds} pattern uses the \texttt{thread\_id\_function} and \texttt{num\_thread\_function} (Figure~\ref{fig:dsl-example}, lines~14--15) to generate the code to perform static work-load distribution; retargeting for a different runtime only requires assigning different values to these variables (e.g., \texttt{ext\_pi\_core\_id} and \texttt{ext\_pi\_cl\_nb\_cores} on PMSIS). 
The \texttt{loop\_body} pattern rebuilds the sequential loop over the computed range. 
 
\item \textbf{OpenMP-level mechanisms.} 
These properties select among alternative realizations of OpenMP mechanisms (e.g., outlining conventions, scheduling policies). 
\rev{Lines~3--4 name the libgomp convention; libomp instead uses \texttt{microtask(ptr\_tid, ptr\_btid, captures)} and \texttt{"by\_pointer"}, which add two runtime-specific thread-identifier parameters and deliver each capture as a separate argument.}
Dynamic scheduling requires specifying two chunk function properties (Figure~\ref{fig:dsl-example}, lines~25--26), called by the dynamic scheduling loop generated by the emit action.
 
\item \textbf{Runtime-specific extensions.} 
These properties are specific to a single runtime.
\texttt{ident} generates a descriptor that carries source-location metadata, which is passed as a pointer to most libomp functions. 
Similarly, \texttt{global\_tid} gets a global thread identifier by calling \texttt{\_\_kmpc\_global\_thread\_num}, and it is also used by a subset of libomp functions.
\end{enumerate}

\subsection[Well-formedness and semantic preservation]{\rev{Well-formedness and semantic preservation}}
\label{sec:preservation}
\rev{Specifications are checked at several points: the parser accepts only well-formed programs, and lowering stops on a construct with no matching declaration or on a plan naming an emit keyword or a property value that no pass implements. Each pass leaves the IR in a form that the MLIR verifier accepts, so any mismatch between a plan and the code it applies to is detected during compilation.
OpenMP semantics (regions, nesting, clauses) are encoded in the \texttt{omp} dialect produced by the ClangIR frontend. The three passes transform the program without altering its semantics: the annotation pass attaches information; the outlining pass relocates a region into a separate function and restructures work-sharing constructs; the plan application pass emits runtime calls and applies structured rewrites.}

\subsection[Scope and limitations]{\rev{Scope and limitations}}
\label{sec:scope}

\rev{Table~\ref{tab:coverage} lists the constructs and clauses supported by our current implementation. PMSIS covers fewer constructs because the bare-metal cluster runtime lacks a tasking API, which is essential to implement constructs such as \texttt{task} and \texttt{taskwait}.  
However, constructs that are not yet supported can still be expressed. For example, constructs like \texttt{critical}, which require a sequence of runtime calls around a specific region, simply need a dedicated declaration, as the existing passes already provide the necessary body-handling mode. Constructs like \texttt{atomic}, which require the lowering to be a single instruction instead of a call, will also need an emit keyword added at that level. 
}

\begin{table*}[t]
\centering
\caption{\rev{Constructs and clauses supported by the current implementation.}}
\label{tab:coverage}
\begin{tabular}{@{}llccc@{}}
\toprule
\textbf{Construct} & \textbf{Supported clauses} &
  \textbf{libomp} & \textbf{libgomp} & \textbf{pmsis} \\
\midrule
\texttt{parallel} & \texttt{num\_threads}, \texttt{proc\_bind},
  \texttt{if}, \texttt{private}, \texttt{firstprivate}
  & \checkmark & \checkmark & \checkmark \\
\texttt{wsloop}   & \texttt{schedule(static)},
  \texttt{schedule(dynamic)}, \texttt{nowait}
  & \checkmark & \checkmark & \checkmark \\
\texttt{barrier}  & \emph{none} & \checkmark & \checkmark & \checkmark \\
\texttt{task}     & \texttt{if}, \texttt{firstprivate}
  & \checkmark & \checkmark & $\times$ \\
\texttt{taskwait} & \emph{none} & \checkmark & \checkmark & $\times$ \\
\bottomrule
\end{tabular}
\end{table*}

\begin{figure*}
\centering
\includegraphics[width=\textwidth]{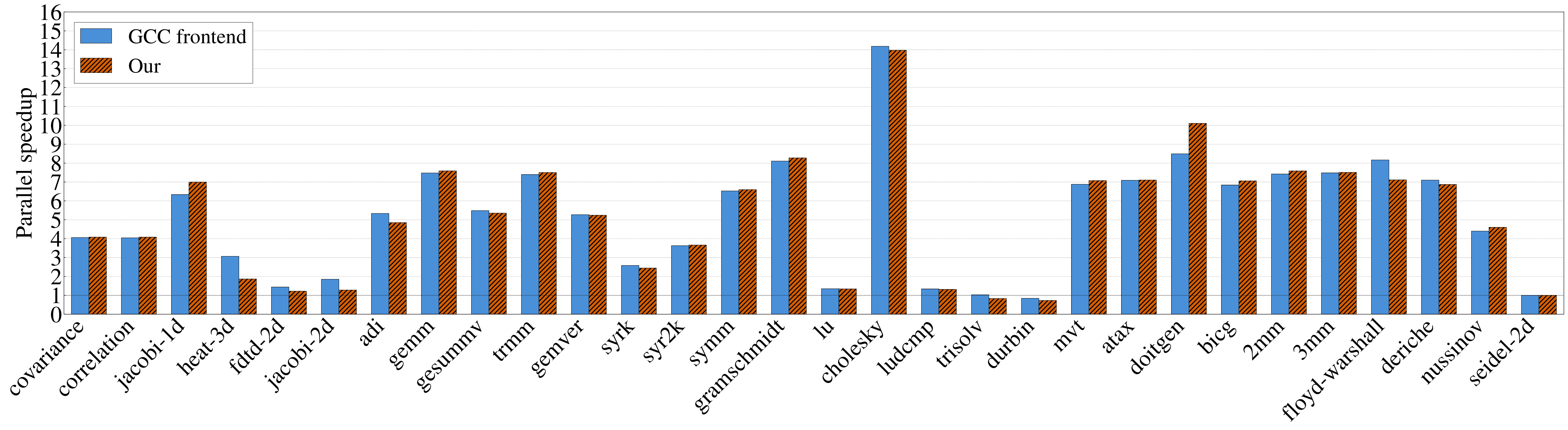}
\caption{Parallel speedup with libgomp: GCC frontend vs our approach (PolyBench/C, large dataset). 
\rev{Each configuration is run ten times with PolyBench's cycle accurate TSC timer; the minimum and maximum are discarded and the mean is reported.}}
\Description{Bar chart comparing parallel speedup across PolyBench benchmarks using GCC frontend and the proposed approach targeting libgomp. Most benchmarks show similar speedups between the two approaches, with slight variations. One benchmark, doitgen, shows higher speedup for the proposed method.}
\label{fig:libgomp_perf}
\end{figure*}

\begin{figure*}
\centering
\includegraphics[width=\textwidth]{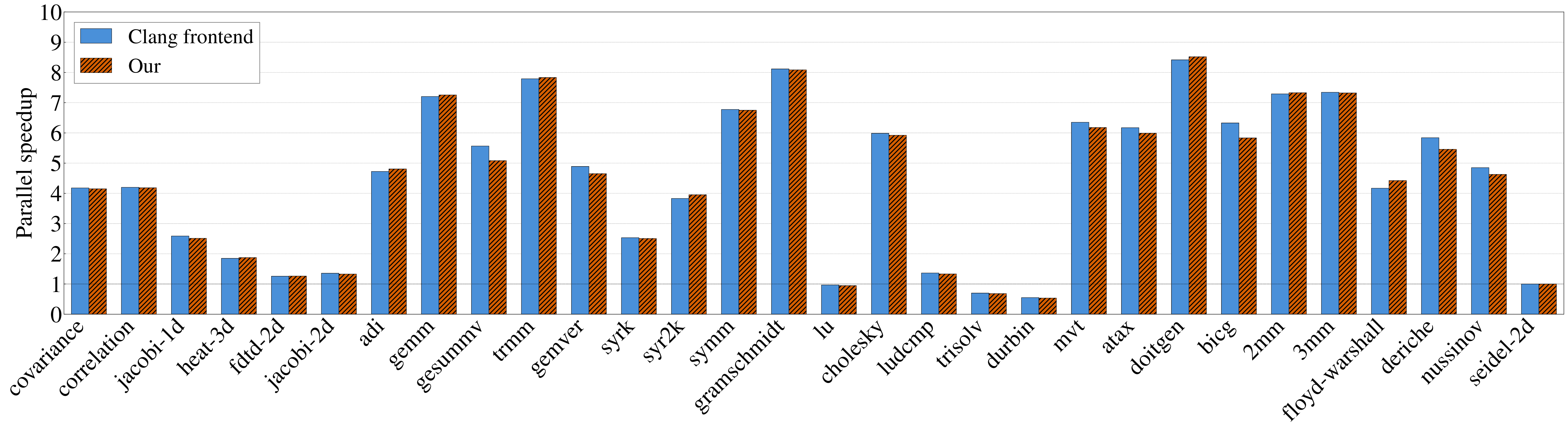}
\caption{Parallel speedup with libomp: Clang frontend vs our approach (PolyBench/C, large dataset).
\rev{Each configuration is run ten times with PolyBench's cycle accurate TSC timer; the minimum and maximum are discarded and the mean is reported.}}
\Description{Bar chart comparing parallel speedup across PolyBench benchmarks using the Clang frontend and the proposed approach targeting libomp. The performance trends are nearly identical across all benchmarks, indicating that the proposed approach preserves performance relative to Clang.}
\label{fig:libomp_perf}
\end{figure*}

\section{Results}
\label{sec:results}
In this section, we assess the effectiveness of our approach using three main criteria: performance portability, engineering effort, and integration with the MLIR optimization pipeline. We compare against state-of-the-art compiler toolchains across both general-purpose and embedded platforms, assessing parallel speedup, code size, and implementation complexity.

\subsection{Experimental setup}
\label{sec:setup}
\paragraph{Benchmark suite.}
We evaluate our approach using the PolyBench/C benchmark suite~\cite{polybench} in its OpenMP-enabled variant (PolyBench/C-OMP). PolyBench is a widely adopted collection of computational kernels representative of numerical workloads, including linear algebra, stencil computations, and dynamic programming.
We selected this benchmark suite for our experimental assessment because it offers a diverse set of computational kernels that cover various domains such as linear algebra, stencil computations, and dynamic programming. These benchmarks are also highly representative of modern numerical and machine-learning workloads.
Their structured, loop-based design makes them especially suitable for evaluating compiler transformations and parallelization strategies in a controlled and reproducible environment.

The suite provides multiple dataset sizes. In our experiments, we use the \texttt{large} configuration on the general-purpose platform to expose sufficient parallelism and stress both computation and memory subsystems. For the PULP platform, we use the \texttt{mini} dataset instead, as this architecture provides limited on-chip memory.

\rev{Tasking is supported on both general-purpose runtimes, where we validated it for functional correctness; the PULP cluster runtime does not provide it. For the performance evaluation, we considered a single benchmark suite executable on every target.}

\paragraph{General-purpose platform.}
Experiments targeting general-purpose runtimes were conducted on an Intel\textsuperscript{\textregistered} Core\texttrademark{} i7-10700K CPU operating at 3.80\,GHz. This processor features 8 physical cores with simultaneous multithreading (SMT), for a total of 16 hardware threads. The system is equipped with 32\,GB of DDR4 main memory.
The memory hierarchy follows a typical multi-level cache organization. Each core has private L1 instruction and data caches (32\,KB each) and a private 256\,KB L2 cache. All cores share a unified last-level cache (L3) of 16\,MB.

\paragraph{PULP platform.}
For the embedded evaluation, we target a GAP8~\cite{gap8} board, which is a physical embodiment of the PULP architecture. The GAP8 system-on-chip integrates a heterogeneous multicore architecture composed of a fabric controller (FC) and a cluster of 8 RISC-V cores dedicated to parallel computation.
The memory hierarchy is explicitly managed with the aim of optimizing energy efficiency. The cluster features a 64\,KB shared L1 scratchpad memory that is software-managed and provides low-latency access for all cores. This memory is backed by a 512\,KB L2 scratchpad shared with the FC. Data transfers between L2 and L1 are orchestrated explicitly via a Direct Memory Access (DMA) engine.

\paragraph{Compiler toolchains.}
On the general-purpose platform, we use two compiler toolchains: GCC 12.3.0 (binary release) and LLVM 23.0.0 (git development branch). The LLVM frontend has been extended to support the lowering of \texttt{omp} dialect operations that are not yet handled by the mainstream CIR pipeline.

For the PULP platform, we evaluated the performance of its custom OpenMP support using the PULP GCC toolchain~\cite{pulp-gcc}, based on GCC 7.1. Although older, this toolchain provides superior performance compared to alternative ones due to backend optimizations tailored to the target.
To evaluate our approach, which is implemented on top of the LLVM infrastructure, we used the PULP LLVM toolchain~\cite{pulp-llvm}. While this toolchain is currently less optimized than its GCC counterpart, we ensure a fair comparison by reporting parallel speedups with respect to the sequential execution produced by the same compiler. This methodology isolates the impact of our lowering strategy from differences in code generation quality caused by compiler backend optimizations.

\paragraph{OpenMP runtime libraries.}
On general-purpose platforms, we evaluate interoperability with standard OpenMP runtimes, including \emph{libgomp} and \emph{libomp}. For PULP, we use the runtime provided by the PULP SDK~\cite{pulp-sdk}, which includes a custom bare-metal OpenMP implementation designed for ultra-low-power embedded systems. This runtime is tightly coupled with the compiler, enabling efficient execution through lightweight barrier synchronization.

\paragraph{Compilation settings.}
All benchmarks are compiled with optimization level \texttt{-O3}. On the CPU platform, OpenMP support is enabled via \texttt{-fopenmp}. For PULP, compilation is performed using the PULP SDK build system. Benchmarks are executed using all available hardware threads (16 on the CPU and 8 on the PULP cluster).

To ensure a fair evaluation of parallelization effects, we turn off automatic vectorization and enforce precise floating-point semantics. These optimizations are orthogonal to our work and could introduce variability that obscures the impact of the lowering strategy on performance.

\subsection{General-purpose runtimes}

We evaluate our approach on general-purpose platforms by comparing against state-of-the-art compiler toolchains targeting standard OpenMP runtimes, namely GCC with \emph{libgomp} and LLVM/Clang with \emph{libomp}.
Figures~\ref{fig:libgomp_perf} and \ref{fig:libomp_perf} report the parallel speedup achieved on the PolyBench/C benchmark suite using all available hardware threads. In the \emph{libgomp} configuration, the baseline corresponds to the GCC frontend, whereas we compare against the LLVM/Clang toolchain in the \emph{libomp} configuration.

Overall, our approach preserves performance across all benchmarks.
The \texttt{doitgen} benchmark exhibits improved performance with our flow when targeting \emph{libgomp}.
This kernel implements a triply-nested matrix update $A[r][q][p] \mathrel{+}= \sum_{s} A[r][q][s] \cdot C4[s][p]$, accumulated through a private buffer \texttt{sum\_local}:

\begin{lstlisting}[language=C]
#pragma omp parallel private(q, p, s)
{
  DATA_TYPE sum_local[NP];
  #pragma omp for
  for (r = 0; r < NR; r++)
    for (q = 0; q < NQ; q++) {
      for (p = 0; p < NP; p++) {
        sum_local[p] = 0.0;
        for (s = 0; s < NP; s++)
          sum_local[p] += A[r][q][s] *
                             C4[s][p];
      }
      for (p = 0; p < NP; p++)
        A[r][q][p] = sum_local[p];
    }
}
\end{lstlisting}

The analysis of the optimized IR produced by the middle-end optimizer reveals that GCC inserts an \texttt{NP~>~0} guard in the first basic block within the region corresponding to the $q$-loop.
This condition is loop-invariant throughout the entire parallel region.
However, GCC fails to hoist it, so the branch executes once per $(r,\,q)$ pair. This overhead is significant in the \texttt{large} dataset configuration, where $NR$ and $NQ$ are 150 and 140, respectively.
Conversely, the LLVM IR optimization flow eliminates this overhead through loop unswitching, because all entry conditions are collapsed into a single short-circuit check before any loop is entered.

\begin{figure*}
\centering
\includegraphics[width=\textwidth]{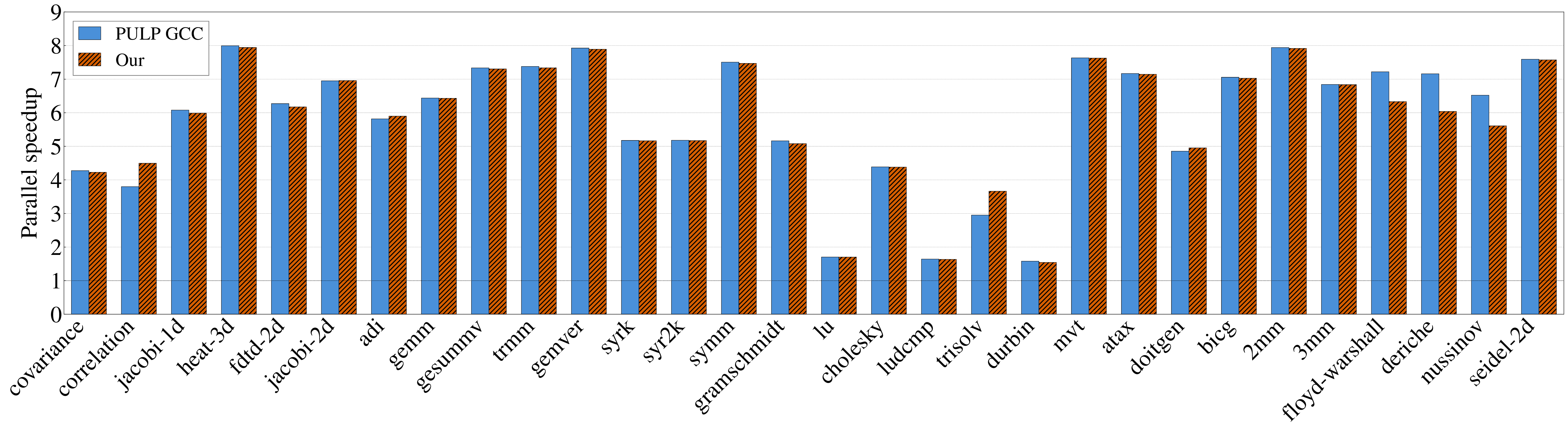}
\caption{Parallel speedup on PULP: GCC (custom libgomp) vs our low-level lowering approach (PolyBench/C, mini dataset).
\rev{Each configuration is a single run: execution on the bare-metal target is deterministic.}}
\Description{Bar chart showing parallel speedup on the PULP platform, comparing the PULP GCC toolchain and the proposed approach. Most benchmarks exhibit similar performance, while a few benchmarks, such as floyd-warshall and deriche, show higher speedups for the GCC baseline.}
\label{fig:pulp_perf}
\end{figure*}

\begin{figure*}
\centering
\includegraphics[width=\textwidth]{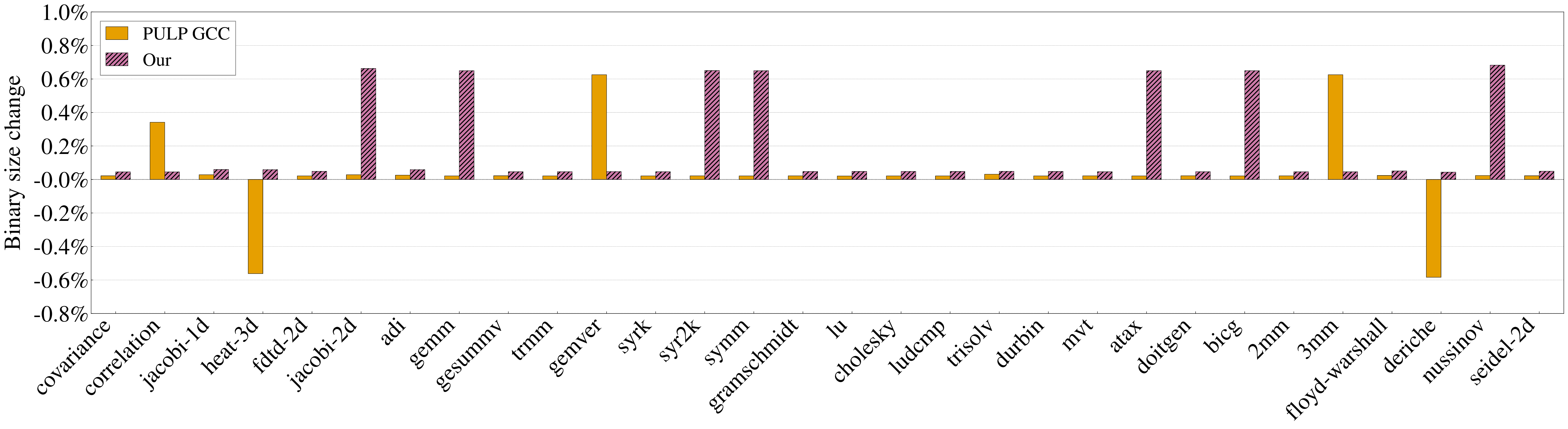}
\caption{Binary size increase for parallel versions on the PULP embedded target.}
\Description{Bar chart showing percentage change in binary size for parallel versions on the PULP platform comparing the proposed approach against the GCC baseline. Most benchmarks show negligible differences, with size increases or decreases remaining within approximately plus or minus one percent.}
\label{fig:pulp_size}
\end{figure*}

\subsection{Bare-metal runtime}
On GAP8, we compared against the PULP GCC toolchain equipped with its custom OpenMP runtime.
Figure~\ref{fig:pulp_perf} reports the parallel speedup obtained across the PolyBench/C benchmark suite. Overall, our approach achieves performance on par with the PULP GCC baseline.

A small subset of benchmarks (i.e., \texttt{floyd-warshall}, \texttt{deriche}, and \texttt{nussinov}) exhibits higher speedups with the PULP GCC toolchain. This behavior is mainly seen in kernels with multiple barrier synchronization points. In these cases, the PULP GCC backend applies target-specific optimizations that reduce the number of instructions required to implement barriers, exploiting knowledge of the underlying hardware and runtime interface.
Incorporating these optimizations into the PULP LLVM toolchain could address this gap, but this goes beyond the scope of this paper.

Figure~\ref{fig:pulp_size} reports the code size of the generated PULP binaries.
We explicitly consider this additional metric for this target since the memory footprint is a critical constraint in ultra-low-power embedded systems. Overall, the differences between the two approaches are negligible across the benchmarks. In a few cases, our approach produces slightly larger binaries; however, the increase remains below $0.7\%$ in all instances.

Overall, these results clearly demonstrate that declarative lowering does not compromise performance or code size even in constrained embedded environments.

\subsection{Lowering Complexity and Engineering Cost}

A central claim of this work is that decoupling OpenMP semantics from runtime-specific implementations not only improves portability but also
significantly reduces the engineering burden to support new targets. In conventional compiler infrastructures, OpenMP lowering is implemented through a large body of tightly coupled code spanning frontend, middle-end, and runtime-specific components. 
As discussed in Section~2.\rev{2}, this design choice hardwires runtime assumptions into the compiler pipeline, making even small extensions or adaptations costly in terms of development effort. To quantify this aspect, we analyze the amount of code required to implement OpenMP lowering in GCC and LLVM, comparing it with our declarative approach.

\sloppypar
For both compilers \rev{we identified the files that participate in OpenMP lowering and counted the lines of the code regions involved, including blank lines and comments. The count covers lowering effort wherever it resides, and excludes directive parsing, AST construction, and semantic checking on all three sides, since our approach reuses those components from the host compiler.}


\begin{table}[htbp]
\centering
\caption{Total and minimal lines of code for OpenMP lowering in GCC,
Clang, and our approach. ``Minimal'' denotes the lines required
to support \texttt{parallel}, \texttt{for}, \texttt{task},
\rev{\texttt{taskwait},} \texttt{barrier},
\texttt{schedule(static|dynamic)} and the data clauses.}
\label{tab:omp-compiler-comparison}
\begin{tabular}{lrr}
\toprule
\textbf{Compiler} & \textbf{Total lines} & \textbf{Minimal lines} \\
\midrule
GCC                   & \rev{82{,}400} & \rev{{\raise0.4ex\hbox{$\scriptstyle\sim$}}21{,}761} \\
Clang                 & \rev{50{,}754} & \rev{{\raise0.4ex\hbox{$\scriptstyle\sim$}}7{,}617} \\
\textbf{Our approach} &             -- & \textbf{\rev{5{,}182}} \\
\bottomrule
\end{tabular}
\end{table}

\rev{In GCC, this spans the dedicated middle-end passes (\texttt{omp-low.cc}, \texttt{omp-expand.cc}, \texttt{omp-general.cc}), a fraction of the gimplification pass (\texttt{gimplify.cc}), and the associated IR definitions and headers.}

To estimate the \emph{minimal} line count, which is the subset required to
support only \texttt{\#pragma omp parallel}, \texttt{\#pragma omp for}, \texttt{schedule(static|dynamic)}, \texttt{\#pragma omp task}, \rev{\texttt{\#pragma omp taskwait},} \texttt{\#pragma omp barrier}, and the data-sharing clauses, we applied a two-stage methodology. First, we identified all top-level functions (or methods) that participate in the lowering of these constructs. 
Then, we classified every function as \emph{minimal} (required by the target feature set) or \emph{non-minimal}. For counting lines, we considered only minimal functions.
For our approach, we reported the total number of lines. All values are reported in Table~\ref{tab:omp-compiler-comparison}.

These results highlight a substantial gap between traditional approaches and our solution.
\rev{In the minimal configuration, the lowering code amounts to $\sim$7,617 lines for Clang and $\sim$21,761 for GCC, whereas our approach requires \rev{5,182} lines for the entire lowering pipeline, including the DSL interpreter and the MLIR passes.}
Overall, this analysis demonstrates that a declarative lowering strategy not only enables performance portability but also significantly lowers the barrier to extending compiler support to new architectures and runtime systems. By transforming lowering from a monolithic, hardcoded process into a programmable abstraction, our approach facilitates rapid experimentation and promotes a more sustainable evolution of compiler infrastructures.

\subsection{Optimizing Before Lowering: Benefits of MLIR Integration}
\label{sec:optimization}
In traditional compilation flows, OpenMP directives are translated early into low-level runtime invocations, obscuring parallel structure and limiting further optimization opportunities. In contrast, our approach maintains OpenMP semantics alongside CIR and other MLIR dialects until the final stages\rev{, so transformations can run before runtime-specific code is generated, enabling general-purpose and OpenMP-specific optimizations.}
\rev{Because DSL-driven lowering rewrites only the \texttt{omp} constructs and the CIR is left untouched, any existing or future CIR transformation applies unchanged; and the \texttt{omp} operations are still in the IR, so transformations that depend on OpenMP structure become possible as well.} 
\rev{After lowering, the resulting LLVM IR still benefits from the standard LLVM optimization pipeline.} 

\begin{figure}
\centering
\includegraphics[width=0.38\textwidth]{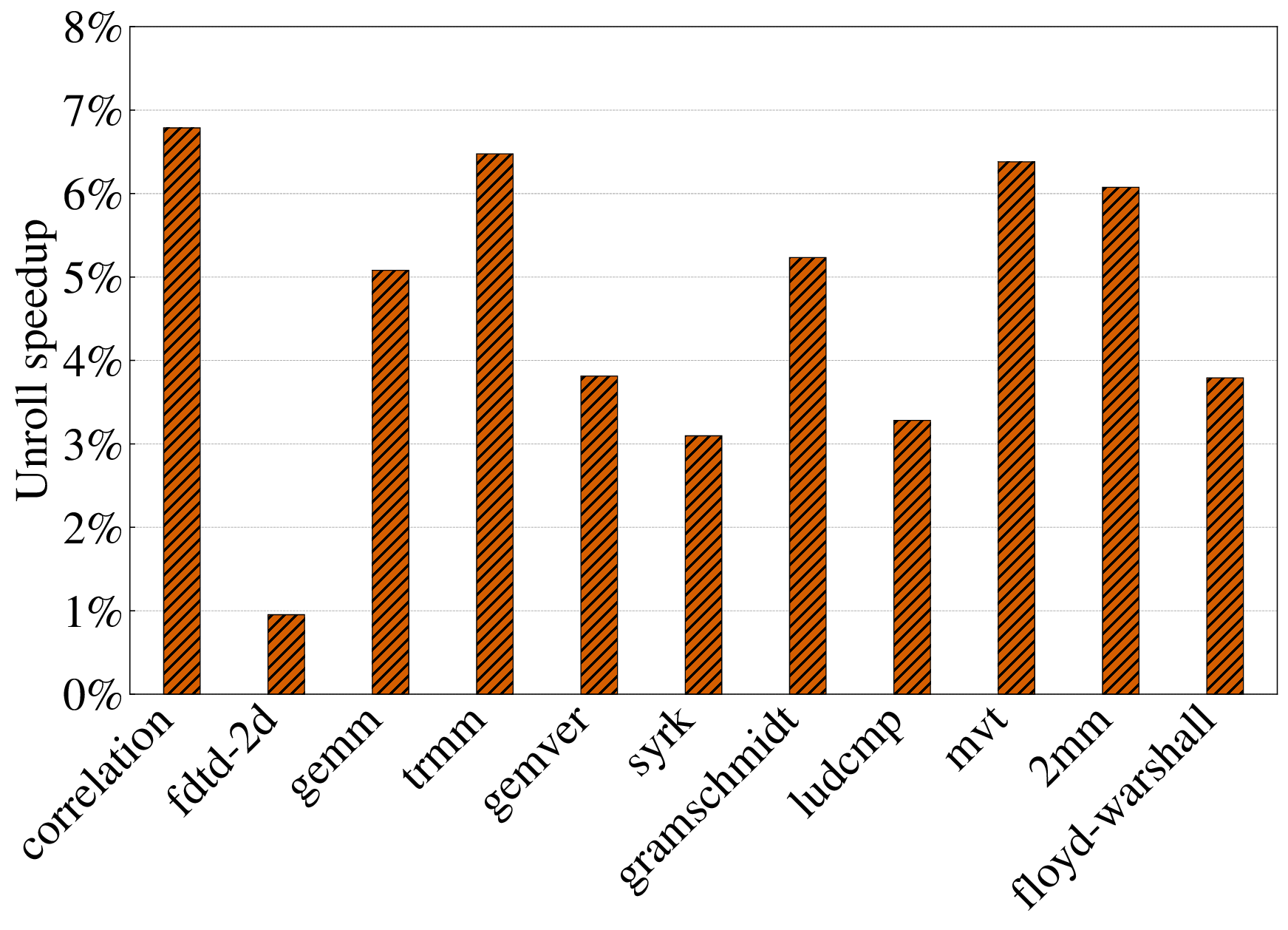}
\caption{Impact of a CIR-level unroll-by-two optimization on PULP parallel speedup.}
\Description{Bar chart illustrating the performance improvement from a loop unrolling optimization applied at the CIR level on the PULP platform. Most benchmarks show speedups around a few percent, while the floyd-warshall benchmark shows a significantly larger improvement of around twenty-four percent.}
\label{fig:pulp-unroll}
\end{figure}

\paragraph{\rev{A general-purpose transformation.}}
\rev{We implemented a custom loop unrolling pass in the CIR dialect. It operates directly on \texttt{cir.for} constructs and unrolls innermost loops by a factor of two, analyzing loop bounds statically and dropping the residual iteration when the trip count is provably even.}
This optimization is particularly effective on the GAP8 platform because it increases instruction-level parallelism and hides the single-cycle latency of the L1 scratchpad, improving overall execution efficiency. Moreover, eliminating redundant condition checks in the inner loops significantly reduces the total execution cycles.

Figure~\ref{fig:pulp-unroll} shows the speedup increment achieved by enabling this optimization. Across ten applications, loop unrolling yields an average improvement of \rev{approximately 4.6\%, ranging from 6.8\% for correlation down to 1.0\% for fdtd-2d}.
The remaining applications in the benchmark suite are only marginally affected by \rev{loop unrolling}.

\paragraph{\rev{An OpenMP-specific transformation.}}
\rev{The implicit barrier that closes a work-sharing loop is redundant when the loop is the last statement of its parallel region, since the following join point already waits for every thread. We implemented a pass on the \texttt{omp} dialect that removes such barriers, along with explicit \texttt{omp.barrier} operations that open a region, close it, or immediately follow another synchronization point}. 
\rev{Traditional compilers apply this elision: Clang only when the loop is written as a combined \texttt{parallel for}, GCC also when the \texttt{parallel} region and the \texttt{for} loop are separate.}
%
%
\rev{Across the PolyBench suite, our pipeline emits 59 team-barrier calls without the pass and 26 with it. On the same kernels, Clang and GCC emit 45 and 28 barriers, respectively. Compared with the previous transformation, the performance impact on the GAP8 platform is minimal: an overall saving of 0.037\%, with 22 of 30 kernels improving and a per-kernel range of $-0.16\%$ to $+1.12\%$ (\texttt{trisolv}). These values are small because the barrier cost on the GAP8 cluster is very low (around 20 cycles) and the kernels are simple, but more complex applications can benefit significantly from this optimization. The few regressions are due to code layout effects rather than the cost of the transformation: removing a call shifts the surrounding code, changing how it maps to the shared instruction cache in the cluster.}


\rev{These experiments highlight} the importance of optimizing before lowering. By leveraging MLIR, our framework enables a flexible compilation strategy where high-level transformations and backend optimizations complement each other.

\section{Related Work}

Prior work relevant to this paper lies at the intersection of OpenMP runtime support, compiler intermediate representations for parallelism, and MLIR-based infrastructures.

\paragraph{OpenMP compiler and runtime implementations.}
A substantial body of research focuses on compiler toolchains and runtime systems that efficiently realize the OpenMP execution model on specific platforms. Mainstream ecosystems such as GCC and LLVM provide mature implementations through tightly integrated runtime libraries, notably \emph{libgomp}~\cite{libgomp} and \emph{libomp}~\cite{libomp}. In these systems, the lowering of OpenMP constructs is tightly coupled with compiler-specific code-generation paths and runtime-specific ABIs. While this design enables high efficiency, it limits portability across runtimes and architectures. This coupling is even more pronounced in embedded and ultra-low-power systems, where runtime support is often highly specialized for the target hardware. Several works~\cite{chapman2009openmp,wang2013libeomp,rossi2014energy} demonstrate OpenMP support on embedded targets through custom implementations, while Agathos et al.~\cite{AGATHOS2022102895} go a step further by proposing a compiler-assisted adaptive runtime organization for embedded multicores. Although these approaches highlight the importance of compiler–runtime co-design, they retain the assumption that lowering is a compiler-internal, target-specific mechanism.

\paragraph{\rev{Decoupling lowering from compiler and runtime.}}
\rev{Earlier research works have loosened that coupling in different ways.
Mercurium~\cite{balart2004mercurium} specifies the transformation of OpenMP directives through code templates, written in the source language.
ROSE~\cite{liao2010rose} translates OpenMP into calls to XOMP, a common runtime library that needs to be implemented on top of each target library.
OpenUH~\cite{liao2007openuh} lowers OpenMP within Open64's IR and can re-emit the result as source code, making the translation portable.
Our approach specifies lowering rules applied directly to the constructs rather than using a template for instantiation or translating to a common library. The calls generated by these rules directly invoke the target runtime, eliminating the need for any additional layers specific to each runtime. Additionally, both general-purpose and OpenMP-aware transformations can be applied to the IR before lowering.
}

\paragraph{Intermediate representations for parallel semantics.}
Another line of work argues that parallel semantics should remain explicit in the compiler intermediate representation, rather than being lowered prematurely into runtime calls. Tapir~\cite{tapir} extends LLVM IR with explicit fork-join constructs, enabling optimization of task-parallel code prior to runtime mapping. Similarly, OpenMP-aware compiler optimizations operate on enriched intermediate representations that preserve OpenMP semantics, allowing analyses and transformations that would otherwise be obscured by early lowering~\cite{doerfert2018openmpopt}. More generally, unified parallel intermediate representations such as UPIR~\cite{wang2023upir} aim to capture parallel semantics across multiple programming models in a common abstraction. While these approaches emphasize preserving parallel structure for optimization and analysis, they focus on representation and transformation rather than on making the lowering itself programmable.

\paragraph{MLIR-based OpenMP compilation and declarative specification.}
MLIR provides a structured OpenMP compilation flow in which directives are represented in the \texttt{omp} dialect and progressively lowered through the MLIR/LLVM stack~\cite{mlir_omp_dialect,mlir_target_llvmir}. While this design preserves high-level semantics and enables composition with other dialects, the final lowering remains tied to a specific runtime ABI through compiler-internal translation logic.

MLIR also provides mechanisms for declaratively specifying transformations. 
DRR~\cite{mlir_drr} and PDLL~\cite{mlir_pdll} express local IR rewrites, while the MLIR Transform dialect~\cite{lucke2025transform} lets performance engineers script fine-grained optimization sequences such as tiling, unrolling, and vectorization over the IR.
%
\rev{These approaches are declarative and IR-native, but domain-agnostic: they focus on \emph{which} transformations are applied to a program rather than on \emph{how} parallel constructs are mapped to a runtime, and they have no vocabulary for OpenMP lowering concepts, so outlining, capture-struct layout, and if-clause serialization fall back to native rewrites.}

Recent work has demonstrated that multi-level IRs can enable retargetable compilation flows and facilitate the integration of domain-specific optimizations across abstraction levels~\cite{lopoukhine2025snitch,fehr2025dialects,fehr2025xdsl}. However, the lowering process itself remains encoded in compiler passes or transformation pipelines, and is not exposed as a programmable abstraction.

\paragraph{Imperative and structure-preserving alternatives.}
A related line of work modularizes OpenMP lowering or preserves high-level program structure, but does so imperatively rather than declaratively. 
The OpenMPIRBuilder, which underlies the MLIR OpenMP translation infrastructure~\cite{mlir_omp_dialect, mlir_target_llvmir}, factors runtime-call emission into helper functions shared across frontends, yet fixes the target runtime in its translation logic: retargeting requires modifying the builder itself.
Our DSL expresses the same mapping declaratively rather than relying on compiler-internal code.
Polygeist~\cite{moses2021polygeist} raises C/C++ into structured MLIR (the \texttt{scf} and \texttt{affine} dialects) to expose programs to polyhedral and affine optimization. It shares the philosophy of preserving high-level structure before lowering, but targets loop-level optimization rather than runtime abstraction: the lowering itself remains a fixed compiler-internal process.

\paragraph{Positioning of this work.}
Our work builds on existing foundations but addresses a different problem. Instead of introducing a new representation or transformation mechanism, we elevate OpenMP lowering to a first-class, declarative, and runtime-agnostic abstraction. 
Unlike previous approaches that either specialize runtime implementations, preserve parallel semantics for optimization, or provide structured compilation infrastructures, our method allows the same OpenMP representation to be reconfigured for different runtimes via a configurable lowering specification, all without modifying the compiler internals.

\section{Conclusion}
This paper introduces a meta-level declarative approach to OpenMP lowering that rethinks the traditional compilation pipeline as a programmable, runtime-agnostic process.
Our framework replaces monolithic lowering with a configurable specification: even with the flexibility of MLIR, the traditional lowering process remains encoded in compiler passes, whereas our approach expresses it through an external DSL.
Our evaluation shows that this flexibility comes at no cost in performance or code size, 
\rev{while the lowering pipeline takes \rev{5,182} lines of code, $\sim$32\% less than the corresponding code in Clang and $\sim$76\% less than in GCC.}

The current implementation supports a subset of OpenMP sufficient for different application domains; supporting additional features, such as offloading and more complex synchronization patterns, is a natural evolution.
More broadly, the declarative lowering paradigm introduced in this paper can be generalized to other programming models (e.g., \rev{SYCL})
and to different scenarios, such as describing the code transformations required to map code to a custom accelerator. This approach would enable a new class of compiler frameworks in which lowering strategies are expressed at a meta-level and adapted to diverse execution contexts.

\begin{acks}
\textit{This work was funded by Infineon Technologies AG.}
\end{acks}


\appendix
\section{Artifact Appendix}

\subsection{Abstract}

This is the supporting artifact for the paper titled ``OpenMP Meta-Lowering: A Declarative Approach to Performance Portable Parallel Code Generation'' as submitted to CGO 2027. It can be used to reproduce all results presented in Section~\ref{sec:results} of this paper.

\subsection{Artifact check-list}

{\small
\begin{itemize}
  \item {\bf Algorithm:} declarative, runtime-parametric lowering of OpenMP constructs: a DSL specification of a runtime ABI is evaluated against the \texttt{omp.*} operations it declares to produce a lowering plan, realized by three MLIR passes (annotation, outlining, plan application), plus an optional redundant team-barrier elimination pass.
  \item {\bf Program:} the PolyBench/C 4.2.1 kernels in their OpenMP-enabled  variant\footnote{\url{https://github.com/EEESlab/PolyBenchC-4.2.1-OpenMP/tree/testing/pulp}} (PolyBench/C-OMP), and hand-written MLIR modules.
  \item {\bf Compilation:} publicly available and included in this artifact: \texttt{mlir-opt-omp}\footnote{\url{https://github.com/EEESlab/mlir-opt-omp/tree/lucap/paper-cleanup}} built against LLVM/MLIR 23.0.0 with ClangIR\footnote{\url{https://github.com/EEESlab/llvm-project/tree/users/lucap/cir-omp-clauses}}, GCC 12.3.0 (Ubuntu binary release), the GAP8 GCC 7.1 toolchain~\footnote{\url{https://github.com/GreenWaves-Technologies/gap-riscv-gnu-toolchain}} and the PULP LLVM 18.1.4 toolchain~\footnote{\url{https://github.com/pulp-platform/llvm-project/tree/integer_SIMD_fix}}.
  \item {\bf Transformations:} the \texttt{mlir-opt-omp} pass pipeline (\texttt{-omp-to-omp-lower}, \texttt{-omp-outline}, \texttt{-omp-lower-plan}, and optionally \texttt{-omp-barrier-elim}), parameterized by \texttt{-omp-lower-dsl} and \texttt{-omp-lower-runtime}.
  \item {\bf Binary:} host ELF executables for the general-purpose experiments and bare-metal RISC-V ELF executables for GAP8, as generated by the workflow in the Docker container image.
  \item {\bf Data set:} inputs generated by PolyBench from the selected dataset macro: \texttt{LARGE\_DATASET} on the general-purpose platform, \texttt{MINI\_DATASET} on GAP8.
  \item {\bf Run-time environment:} the Docker container image platform is linux/amd64. Host runs use the OpenMP runtime shipped with each toolchain (\emph{libgomp} and \emph{libomp}); GAP8 runs use the OpenMP support of the PULP SDK.
  \item {\bf Hardware:} an x86-64 multicore machine for the general-purpose experiments, of the class described in Section~\ref{sec:setup}. The GAP8 results are obtained on the cycle-accurate \texttt{gvsoc} simulator shipped in the image.
  \item {\bf Run-time state:} other significant loads can impact general-purpose timings; GAP8 simulated runs are unaffected.
  %
  %
  \item {\bf Metrics:} parallel speedup against the sequential execution
    produced by the same compiler, binary size, static count of emitted team
    barriers, and lines of code.
  \item {\bf Output:} execution logs and CSV files, along with the derived
    plots and tables of Section~\ref{sec:results}.
  \item {\bf Experiments:} reproduction following the
    instructions in this appendix and using the scripts available in the artifact.
  \item {\bf How much disk space required:} 20.6\,GB for the
    Docker container image, and less than a GB more for the results of the full benchmark suite.
  \item {\bf How much time is needed to prepare workflow:}
    about 15 minutes, i.e.\ the time to download and load the container image; everything it needs is already built inside it.
  \item {\bf How much time is needed to complete experiments:} on the platform of Section~\ref{sec:setup}, the two
    general-purpose performance runs (\emph{libomp} and \emph{libgomp}, ten
    repetitions of every timing cell) take about 5 hours together; every other
    experiment adds less than 30 minutes. Reducing the repetition count to
    \texttt{REPS=1} brings the complete artifact under one hour, at the cost of
    any measure of run-to-run variation
    (Section~\ref{sec:ae-customization}).

  \item {\bf Publicly available:} Yes.
  \item {\bf Code licenses:} Apache License version
    2.0 with LLVM Exceptions.
  \item {\bf Data licenses:} PolyBench/C 4.2.1 and its
    OpenMP variant, under the Ohio State University Software Distribution License.
  \item {\bf Workflow framework used:} none.
  \item {\bf Archived:} \url{https://doi.org/10.5281/zenodo.22211483} (Zenodo DOI).
\end{itemize}
}

\subsection{Description}

\subsubsection{How delivered}

The artifact is archived at \url{https://doi.org/10.5281/zenodo.22211483} (Zenodo DOI) as a Docker container image
containing all the toolchains, benchmarks, and scripts needed to reproduce the results.

\subsubsection{Hardware dependencies}

An x86-64 machine running Linux, with at least 20.6\,GB of free disk space. The general-purpose experiments report parallel speedups, so the core count of the
host bounds what can be observed: reproducing Figures~\ref{fig:libgomp_perf} and~\ref{fig:libomp_perf} as published requires a machine of comparable core count to the one of Section~\ref{sec:setup}, whereas the shape of the comparison
between the baseline and our approach is preserved on any multicore host. The GAP8 experiments are simulated and have no hardware dependency.

\subsubsection{Software dependencies}

A working Docker installation. This has been tested on a Linux x86-64 host machine with Docker 29.1.3. Internet access is needed only to download the image and count lines of code for upstream GCC and LLVM versions.

\subsection{Installation}

Download the Docker image from the archive, navigate to the directory
containing the download, and load it:

{\small
\begin{verbatim}
$ docker image load --input \
  mlir-opt-omp-1.0.0-linux-amd64.docker.tar.zst
\end{verbatim}
}

\noindent Check that the container starts:

{\small
\begin{verbatim}
$ docker run --rm -it mlir-opt-omp:latest
\end{verbatim}
}

\noindent Check the availability of the compilers:
{\small
\begin{verbatim}
$ docker run --rm mlir-opt-omp:latest \
  bash -c "gcc-12 --version"
\end{verbatim}
}
{\small
\begin{verbatim}
$ docker run --rm mlir-opt-omp:latest \
  bash -c "/opt/llvm-mlir/bin/clang --version"
\end{verbatim}
}
{\small
\begin{verbatim}
$ docker run --rm mlir-opt-omp:latest \
  bash -c "riscv32-unknown-elf-gcc --version"
\end{verbatim}
}
{\small
\begin{verbatim}
$ docker run --rm mlir-opt-omp:latest \
  bash -c "/opt/pulp-llvm/bin/clang --version"
\end{verbatim}
}

\subsection{Experiment workflow}

\noindent The commands below must be executed in the \texttt{/root/mlir-opt-omp} folder within the Docker environment (the default folder at startup).

\paragraph{E1 --- Passes.} 74 lit tests over the three passes:

{\small
\begin{verbatim}
$ cd ~/mlir-opt-omp
$ cmake --build build --target check-omp
\end{verbatim}
}

\paragraph{E2 --- Correctness.} Array dumps compared bit-by-bit against the reference toolchain, 30 kernels per runtime:

{\small
\begin{verbatim}
$ cd ~/mlir-opt-omp/test/Integration
$ RUNTIME=iomp ./run_correctness.sh
$ RUNTIME=libgomp ./run_correctness.sh
$ RUNTIME=pmsis ./run_correctness.sh
\end{verbatim}
}

\paragraph{E3 --- Performance.} One configuration per figure:

{\small
\begin{verbatim}
$ cd ~/mlir-opt-omp/test/Integration
$ RUN_ENV=configs/paper-libgomp.env \
    PLOT=true ./run_performance.sh
$ RUN_ENV=configs/paper-iomp.env \
    PLOT=true ./run_performance.sh
$ RUN_ENV=configs/paper-pmsis.env \
    PLOT=true ./run_performance.sh
\end{verbatim}
}

\noindent The first two generate Figures~\ref{fig:libgomp_perf} and~\ref{fig:libomp_perf}, about 5 hours together; the third generates Figures~\ref{fig:pulp_perf} and~\ref{fig:pulp_size}.

\paragraph{E4 --- Section~\ref{sec:optimization}.} The results of the proposed transformations:

{\small
\begin{verbatim}
$ cd ~/mlir-opt-omp/test/Integration
$ ./run_unroll.sh              # unroll Figure
$ ./run_barrier_vs_native.sh   # barrier counts
$ RUNTIME=pmsis BARRIER_ELIM=both \
    ./run_performance.sh       # compare run cycles
\end{verbatim}
}
The first generates Figure~\ref{fig:pulp-unroll}; the last compares cycles with and without barrier elimination.

\paragraph{E5 --- Table~\ref{tab:omp-compiler-comparison}.} Line counts. The GCC and LLVM trees should be cloned at the commits pinned in \texttt{test/LoC/README.md} before running the python scripts:

{\small
\begin{verbatim}
$ cd ~/mlir-opt-omp
$ python3 mlir-opt-omp_loc.py
$ python3 test/LoC/gcc_omp_loc.py --root $GCC_SRC
$ python3 test/LoC/clang_omp_loc.py \
    --root $LLVM_SRC
\end{verbatim}
}

\paragraph{E6 --- Table~\ref{tab:coverage}.} Twelve MLIR programs covering the
constructs and clauses that never appear in the PolyBench suite.

{\small
\begin{verbatim}
$ cd ~/mlir-opt-omp/test/Integration/constructs
$ RUNTIME=iomp    ./run_constructs.sh
$ RUNTIME=libgomp ./run_constructs.sh
\end{verbatim}
}

\subsection{Evaluation and expected result}

E1 expects 74 passing tests; E2 expects \texttt{PASS} for every kernel; E5 expects the line counts of Table~2; E4 barrier expects the count reported at Section~\ref{sec:optimization}; E6 expects \texttt{PASS} on all twelve tests.

The GAP8 runs (E3 and E4) are simulated on a
bare-metal, in-order cluster, so they are deterministic and are expected to closely replicate Figures~\ref{fig:pulp_perf}, \ref{fig:pulp_size}, and~\ref{fig:pulp-unroll}.

The host speedups in Figures~\ref{fig:libgomp_perf} and~\ref{fig:libomp_perf} scale with the machine, so what is evaluated is not the absolute speedup but the per-kernel ratio between ours and the native toolchain, and its geometric mean across the suite.

\subsection{Experiment customization}
\label{sec:ae-customization}

Every script reads \texttt{RUNTIME}, \texttt{DATASET}, \texttt{THREADS}, \texttt{KERNELS}, \texttt{REPS}, \texttt{BARRIER\_ELIM}, \texttt{PLOT} and \texttt{OUTDIR}, from the command line or from a \texttt{RUN\_ENV} file. Each runtime and the barrier-elimination variant write to their own directories.

For a shorter run, it is always possible to reduce the number of \texttt{REPS}.

\subsection{Notes}

\begin{itemize}
  \item Vectorization off and precise floating-point semantics (Section~\ref{sec:setup}) are what make the bit-exact correctness oracle possible: with contraction and reassociation enabled, two binaries diverge on the iterative kernels due to IEEE-754 floating-point arithmetic.
  \item The runtime identified as \emph{libomp} in the paper is referenced as \texttt{iomp} throughout the
    artifact: in \texttt{rules.dsl}, in \texttt{-omp-lower-runtime}, and in the names of the result directories. \texttt{outline\_signature} has been removed since it was redundant: the signature of the outlined function follows from  \texttt{capture\_strategy} (\texttt{by\_pointer}, \texttt{packed}, \texttt{shareds}), as no evaluated runtime needs the two to vary independently.
    Clauses whose value is known only at run time (e.g., the \texttt{if} clause on \texttt{parallel} and \texttt{task}) are handled by a \texttt{branch} block, which a \texttt{when} guard cannot express.
    Dynamic scheduling is expressed by \texttt{first\_chunk}/\texttt{next\_chunk} blocks in place of the \texttt{chunk\_*\_function} properties: a block carries the call itself, arguments included, where the property gave only the name of the function to call.
\end{itemize}


\bibliographystyle{ACM-Reference-Format}
\bibliography{references}


\begin{thebibliography}{35}


\ifx \showCODEN    \undefined \def \showCODEN     #1{\unskip}     \fi
\ifx \showDOI      \undefined \def \showDOI       #1{#1}\fi
\ifx \showISBNx    \undefined \def \showISBNx     #1{\unskip}     \fi
\ifx \showISBNxiii \undefined \def \showISBNxiii  #1{\unskip}     \fi
\ifx \showISSN     \undefined \def \showISSN      #1{\unskip}     \fi
\ifx \showLCCN     \undefined \def \showLCCN      #1{\unskip}     \fi
\ifx \shownote     \undefined \def \shownote      #1{#1}          \fi
\ifx \showarticletitle \undefined \def \showarticletitle #1{#1}   \fi
\ifx \showURL      \undefined \def \showURL       {\relax}        \fi
\providecommand\bibfield[2]{#2}
\providecommand\bibinfo[2]{#2}
\providecommand\natexlab[1]{#1}
\providecommand\showeprint[2][]{arXiv:#2}

\bibitem[Agathos et~al\mbox{.}(2022)]%
        {AGATHOS2022102895}
\bibfield{author}{\bibinfo{person}{Spiros~N. Agathos}, \bibinfo{person}{Vassilios~V. Dimakopoulos}, {and} \bibinfo{person}{Ilias~K. Kasmeridis}.} \bibinfo{year}{2022}\natexlab{}.
\newblock \showarticletitle{Compiler-assisted, adaptive runtime system for the support of OpenMP in embedded multicores}.
\newblock \bibinfo{journal}{\emph{Parallel Comput.}}  \bibinfo{volume}{110} (\bibinfo{year}{2022}), \bibinfo{pages}{102895}.
\newblock
\showISSN{0167-8191}
\urldef\tempurl%
\url{https://doi.org/10.1016/j.parco.2022.102895}
\showDOI{\tempurl}


\bibitem[Balart et~al\mbox{.}(2004)]%
        {balart2004mercurium}
\bibfield{author}{\bibinfo{person}{Jairo Balart}, \bibinfo{person}{Alejandro Duran}, \bibinfo{person}{Marc Gonz{\`a}lez}, \bibinfo{person}{Xavier Martorell}, \bibinfo{person}{Eduard Ayguad{\'e}}, {and} \bibinfo{person}{Jes{\'u}s Labarta}.} \bibinfo{year}{2004}\natexlab{}.
\newblock \showarticletitle{Nanos Mercurium: A Research Compiler for {OpenMP}}. In \bibinfo{booktitle}{\emph{Proceedings of the European Workshop on OpenMP (EWOMP)}}. \bibinfo{pages}{103--109}.
\newblock


\bibitem[Chapman et~al\mbox{.}(2009)]%
        {chapman2009openmp}
\bibfield{author}{\bibinfo{person}{Barbara Chapman}, \bibinfo{person}{Lei Huang}, \bibinfo{person}{Eric Biscondi}, \bibinfo{person}{Eric Stotzer}, \bibinfo{person}{Ashish Shrivastava}, {and} \bibinfo{person}{Alan Gatherer}.} \bibinfo{year}{2009}\natexlab{}.
\newblock \showarticletitle{Implementing OpenMP on a high performance embedded multicore MPSoC}. In \bibinfo{booktitle}{\emph{2009 IEEE International Symposium on Parallel \& Distributed Processing}}. \bibinfo{publisher}{IEEE}, \bibinfo{address}{Rome, Italy}, \bibinfo{pages}{1--8}.
\newblock
\urldef\tempurl%
\url{https://doi.org/10.1109/IPDPS.2009.5161107}
\showDOI{\tempurl}


\bibitem[Conti et~al\mbox{.}(2025)]%
        {conti2025pulp}
\bibfield{author}{\bibinfo{person}{Francesco Conti}, \bibinfo{person}{Angelo Garofalo}, \bibinfo{person}{Davide Rossi}, \bibinfo{person}{Giuseppe Tagliavini}, {and} \bibinfo{person}{Luca Benini}.} \bibinfo{year}{2025}\natexlab{}.
\newblock \showarticletitle{Open Source Heterogeneous SoCs for Artificial Intelligence: The PULP Platform experience}.
\newblock \bibinfo{journal}{\emph{IEEE Solid-State Circuits Magazine}} \bibinfo{volume}{17}, \bibinfo{number}{2} (\bibinfo{year}{2025}), \bibinfo{pages}{49--60}.
\newblock
\urldef\tempurl%
\url{https://doi.org/10.1109/MSSC.2025.3537987}
\showDOI{\tempurl}


\bibitem[Dagum and Menon(1998)]%
        {dagum1998openmp}
\bibfield{author}{\bibinfo{person}{L. Dagum} {and} \bibinfo{person}{R. Menon}.} \bibinfo{year}{1998}\natexlab{}.
\newblock \showarticletitle{OpenMP: an industry standard API for shared-memory programming}.
\newblock \bibinfo{journal}{\emph{IEEE Computational Science and Engineering}} \bibinfo{volume}{5}, \bibinfo{number}{1} (\bibinfo{year}{1998}), \bibinfo{pages}{46--55}.
\newblock
\urldef\tempurl%
\url{https://doi.org/10.1109/99.660313}
\showDOI{\tempurl}


\bibitem[Deakin et~al\mbox{.}(2019)]%
        {deakin2019perfportability}
\bibfield{author}{\bibinfo{person}{Tom Deakin}, \bibinfo{person}{Simon McIntosh-Smith}, \bibinfo{person}{James Price}, \bibinfo{person}{Andrei Poenaru}, \bibinfo{person}{Patrick Atkinson}, \bibinfo{person}{Codrin Popa}, {and} \bibinfo{person}{Justin Salmon}.} \bibinfo{year}{2019}\natexlab{}.
\newblock \showarticletitle{Performance Portability across Diverse Computer Architectures}. In \bibinfo{booktitle}{\emph{2019 IEEE/ACM International Workshop on Performance, Portability and Productivity in HPC (P3HPC)}}. \bibinfo{publisher}{IEEE}, \bibinfo{address}{Denver, CO, USA}, \bibinfo{pages}{1--13}.
\newblock
\urldef\tempurl%
\url{https://doi.org/10.1109/P3HPC49587.2019.00006}
\showDOI{\tempurl}


\bibitem[Diaz et~al\mbox{.}(2012)]%
        {diaz2012survey}
\bibfield{author}{\bibinfo{person}{Javier Diaz}, \bibinfo{person}{Camelia Muñoz-Caro}, {and} \bibinfo{person}{Alfonso Niño}.} \bibinfo{year}{2012}\natexlab{}.
\newblock \showarticletitle{A Survey of Parallel Programming Models and Tools in the Multi and Many-Core Era}.
\newblock \bibinfo{journal}{\emph{IEEE Transactions on Parallel and Distributed Systems}} \bibinfo{volume}{23}, \bibinfo{number}{8} (\bibinfo{year}{2012}), \bibinfo{pages}{1369--1386}.
\newblock
\urldef\tempurl%
\url{https://doi.org/10.1109/TPDS.2011.308}
\showDOI{\tempurl}


\bibitem[Doerfert and Finkel(2018)]%
        {doerfert2018openmpopt}
\bibfield{author}{\bibinfo{person}{Johannes Doerfert} {and} \bibinfo{person}{Hal Finkel}.} \bibinfo{year}{2018}\natexlab{}.
\newblock \showarticletitle{Compiler Optimizations for OpenMP}. In \bibinfo{booktitle}{\emph{Proceedings of the International Workshop on OpenMP (IWOMP)}}. \bibinfo{publisher}{Springer}, \bibinfo{address}{Barcelona, Spain}, \bibinfo{pages}{113--127}.
\newblock
\urldef\tempurl%
\url{https://doi.org/10.1007/978-3-319-98521-3_8}
\showDOI{\tempurl}


\bibitem[Fehr et~al\mbox{.}(2025a)]%
        {fehr2025dialects}
\bibfield{author}{\bibinfo{person}{Mathieu Fehr}, \bibinfo{person}{Yuyou Fan}, \bibinfo{person}{Hugo Pompougnac}, \bibinfo{person}{John Regehr}, {and} \bibinfo{person}{Tobias Grosser}.} \bibinfo{year}{2025}\natexlab{a}.
\newblock \showarticletitle{First-Class Verification Dialects for MLIR}.
\newblock \bibinfo{journal}{\emph{Proceedings of the ACM on Programming Languages}} \bibinfo{volume}{9}, \bibinfo{number}{PLDI}, Article \bibinfo{articleno}{206} (\bibinfo{date}{June} \bibinfo{year}{2025}), \bibinfo{numpages}{25}~pages.
\newblock
\urldef\tempurl%
\url{https://doi.org/10.1145/3729309}
\showDOI{\tempurl}


\bibitem[Fehr et~al\mbox{.}(2025b)]%
        {fehr2025xdsl}
\bibfield{author}{\bibinfo{person}{Mathieu Fehr}, \bibinfo{person}{Michel Weber}, \bibinfo{person}{Christian Ulmann}, \bibinfo{person}{Alexandre Lopoukhine}, \bibinfo{person}{Martin~Paul L\"{u}cke}, \bibinfo{person}{Th\'{e}o Degioanni}, \bibinfo{person}{Christos Vasiladiotis}, \bibinfo{person}{Michel Steuwer}, {and} \bibinfo{person}{Tobias Grosser}.} \bibinfo{year}{2025}\natexlab{b}.
\newblock \showarticletitle{xDSL: Sidekick Compilation for SSA-Based Compilers}. In \bibinfo{booktitle}{\emph{Proceedings of the 23rd ACM/IEEE International Symposium on Code Generation and Optimization}} (Las Vegas, NV, USA) \emph{(\bibinfo{series}{CGO '25})}. \bibinfo{publisher}{Association for Computing Machinery}, \bibinfo{address}{New York, NY, USA}, \bibinfo{pages}{179–192}.
\newblock
\urldef\tempurl%
\url{https://doi.org/10.1145/3696443.3708945}
\showDOI{\tempurl}


\bibitem[Flamand et~al\mbox{.}(2018)]%
        {gap8}
\bibfield{author}{\bibinfo{person}{Eric Flamand}, \bibinfo{person}{Davide Rossi}, \bibinfo{person}{Francesco Conti}, \bibinfo{person}{Igor Loi}, \bibinfo{person}{Antonio Pullini}, \bibinfo{person}{Florent Rotenberg}, {and} \bibinfo{person}{Luca Benini}.} \bibinfo{year}{2018}\natexlab{}.
\newblock \showarticletitle{GAP-8: A RISC-V SoC for AI at the Edge of the IoT}. In \bibinfo{booktitle}{\emph{2018 IEEE 29th International Conference on Application-specific Systems, Architectures and Processors (ASAP)}}. \bibinfo{publisher}{IEEE Press}, \bibinfo{address}{Piscataway, NJ, USA}, \bibinfo{pages}{1--4}.
\newblock
\urldef\tempurl%
\url{https://doi.org/10.1109/ASAP.2018.8445101}
\showDOI{\tempurl}


\bibitem[{GNU Project}(2026)]%
        {libgomp}
\bibfield{author}{\bibinfo{person}{{GNU Project}}.} \bibinfo{year}{2026}\natexlab{}.
\newblock \bibinfo{title}{GNU OpenMP Runtime Library (libgomp)}.
\newblock \bibinfo{howpublished}{\url{https://gcc.gnu.org/onlinedocs/libgomp/}}.
\newblock
\newblock
\shownote{Accessed: 2026-04-01}.


\bibitem[Kadosh et~al\mbox{.}(2023)]%
        {kadosh2023openmp}
\bibfield{author}{\bibinfo{person}{Tal Kadosh}, \bibinfo{person}{Niranjan Hasabnis}, \bibinfo{person}{Timothy Mattson}, \bibinfo{person}{Yuval Pinter}, {and} \bibinfo{person}{Gal Oren}.} \bibinfo{year}{2023}\natexlab{}.
\newblock \showarticletitle{Quantifying OpenMP: Statistical Insights into Usage and Adoption}. In \bibinfo{booktitle}{\emph{2023 IEEE High Performance Extreme Computing Conference (HPEC)}}. \bibinfo{publisher}{IEEE}, \bibinfo{address}{Boston, MA, USA}, \bibinfo{pages}{1--7}.
\newblock
\urldef\tempurl%
\url{https://doi.org/10.1109/HPEC58863.2023.10363459}
\showDOI{\tempurl}


\bibitem[Lattner et~al\mbox{.}(2020)]%
        {lattner2020mlir}
\bibfield{author}{\bibinfo{person}{Chris Lattner}, \bibinfo{person}{Mehdi Amini}, \bibinfo{person}{Uday Bondhugula}, \bibinfo{person}{Albert Cohen}, \bibinfo{person}{Andy Davis}, \bibinfo{person}{Jacques Pienaar}, \bibinfo{person}{River Riddle}, \bibinfo{person}{Tatiana Shpeisman}, \bibinfo{person}{Nicolas Vasilache}, {and} \bibinfo{person}{Oleksandr Zinenko}.} \bibinfo{year}{2020}\natexlab{}.
\newblock \bibinfo{title}{{MLIR}: A Compiler Infrastructure for the End of {Moore's} Law}.
\newblock
\newblock
\urldef\tempurl%
\url{https://doi.org/10.48550/arXiv.2002.11054}
\showDOI{\tempurl}


\bibitem[Lattner et~al\mbox{.}(2021)]%
        {lattner2021mlir}
\bibfield{author}{\bibinfo{person}{Chris Lattner}, \bibinfo{person}{Mehdi Amini}, \bibinfo{person}{Uday Bondhugula}, \bibinfo{person}{Albert Cohen}, \bibinfo{person}{Andy Davis}, \bibinfo{person}{Jacques Pienaar}, \bibinfo{person}{River Riddle}, \bibinfo{person}{Tatiana Shpeisman}, \bibinfo{person}{Nicolas Vasilache}, {and} \bibinfo{person}{Oleksandr Zinenko}.} \bibinfo{year}{2021}\natexlab{}.
\newblock \showarticletitle{MLIR: scaling compiler infrastructure for domain specific computation}. In \bibinfo{booktitle}{\emph{Proceedings of the 2021 IEEE/ACM International Symposium on Code Generation and Optimization}} (Virtual Event, Republic of Korea) \emph{(\bibinfo{series}{CGO '21})}. \bibinfo{publisher}{IEEE Press}, \bibinfo{address}{Seoul, Republic of Korea}, \bibinfo{pages}{2–14}.
\newblock
\urldef\tempurl%
\url{https://doi.org/10.1109/CGO51591.2021.9370308}
\showDOI{\tempurl}


\bibitem[Liao et~al\mbox{.}(2007)]%
        {liao2007openuh}
\bibfield{author}{\bibinfo{person}{Chunhua Liao}, \bibinfo{person}{Oscar Hernandez}, \bibinfo{person}{Barbara Chapman}, \bibinfo{person}{Wenguang Chen}, {and} \bibinfo{person}{Weimin Zheng}.} \bibinfo{year}{2007}\natexlab{}.
\newblock \showarticletitle{{OpenUH}: An Optimizing, Portable {OpenMP} Compiler}.
\newblock \bibinfo{journal}{\emph{Concurrency and Computation: Practice and Experience}} \bibinfo{volume}{19}, \bibinfo{number}{18} (\bibinfo{year}{2007}), \bibinfo{pages}{2317--2332}.
\newblock
\urldef\tempurl%
\url{https://doi.org/10.1002/cpe.1174}
\showDOI{\tempurl}


\bibitem[Liao et~al\mbox{.}(2010)]%
        {liao2010rose}
\bibfield{author}{\bibinfo{person}{Chunhua Liao}, \bibinfo{person}{Daniel~J. Quinlan}, \bibinfo{person}{Thomas Panas}, {and} \bibinfo{person}{Bronis~R. de Supinski}.} \bibinfo{year}{2010}\natexlab{}.
\newblock \showarticletitle{A {ROSE}-Based {OpenMP} 3.0 Research Compiler Supporting Multiple Runtime Libraries}. In \bibinfo{booktitle}{\emph{Proceedings of the 6th International Workshop on OpenMP (IWOMP)}} \emph{(\bibinfo{series}{Lecture Notes in Computer Science}, Vol.~\bibinfo{volume}{6132})}. \bibinfo{publisher}{Springer}, \bibinfo{address}{Berlin, Heidelberg}, \bibinfo{pages}{15--28}.
\newblock
\urldef\tempurl%
\url{https://doi.org/10.1007/978-3-642-13217-9_2}
\showDOI{\tempurl}


\bibitem[{LLVM Project}(2024a)]%
        {mlir_target_llvmir}
\bibfield{author}{\bibinfo{person}{{LLVM Project}}.} \bibinfo{year}{2024}\natexlab{a}.
\newblock \bibinfo{title}{{LLVM IR Target} -- {MLIR}}.
\newblock \bibinfo{howpublished}{\url{https://mlir.llvm.org/docs/TargetLLVMIR/}}.
\newblock
\newblock
\shownote{Accessed: 2026-04-01}.


\bibitem[{LLVM Project}(2024b)]%
        {mlir_pdll}
\bibfield{author}{\bibinfo{person}{{LLVM Project}}.} \bibinfo{year}{2024}\natexlab{b}.
\newblock \bibinfo{title}{PDLL - PDL Language - MLIR}.
\newblock \bibinfo{howpublished}{\url{https://mlir.llvm.org/docs/PDLL/}}.
\newblock
\newblock
\shownote{Accessed: 2026-04-01}.


\bibitem[{LLVM Project}(2024c)]%
        {mlir_drr}
\bibfield{author}{\bibinfo{person}{{LLVM Project}}.} \bibinfo{year}{2024}\natexlab{c}.
\newblock \bibinfo{title}{Table-driven Declarative Rewrite Rule (DRR) - MLIR}.
\newblock \bibinfo{howpublished}{\url{https://mlir.llvm.org/docs/DeclarativeRewrites/}}.
\newblock
\newblock
\shownote{Accessed: 2026-04-01}.


\bibitem[{LLVM Project}(2024d)]%
        {mlir_omp_dialect}
\bibfield{author}{\bibinfo{person}{{LLVM Project}}.} \bibinfo{year}{2024}\natexlab{d}.
\newblock \bibinfo{title}{\texttt{omp} {Dialect} -- {MLIR}}.
\newblock \bibinfo{howpublished}{\url{https://mlir.llvm.org/docs/Dialects/OpenMPDialect/}}.
\newblock
\newblock
\shownote{Accessed: 2026-04-01}.


\bibitem[{LLVM Project}(2026a)]%
        {clangir}
\bibfield{author}{\bibinfo{person}{{LLVM Project}}.} \bibinfo{year}{2026}\natexlab{a}.
\newblock \bibinfo{title}{ClangIR}.
\newblock
\newblock
\urldef\tempurl%
\url{https://llvm.github.io/clangir/}
\showURL{%
\tempurl}
\newblock
\shownote{Accessed: 2026-04-01}.


\bibitem[{LLVM Project}(2026b)]%
        {libomp}
\bibfield{author}{\bibinfo{person}{{LLVM Project}}.} \bibinfo{year}{2026}\natexlab{b}.
\newblock \bibinfo{title}{LLVM OpenMP Runtime (libomp)}.
\newblock \bibinfo{howpublished}{\url{https://openmp.llvm.org/}}.
\newblock
\newblock
\shownote{Accessed: 2026-04-01}.


\bibitem[Lopoukhine et~al\mbox{.}(2025)]%
        {lopoukhine2025snitch}
\bibfield{author}{\bibinfo{person}{Alexandre Lopoukhine}, \bibinfo{person}{Federico Ficarelli}, \bibinfo{person}{Christos Vasiladiotis}, \bibinfo{person}{Anton Lydike}, \bibinfo{person}{Josse Van~Delm}, \bibinfo{person}{Alban Dutilleul}, \bibinfo{person}{Luca Benini}, \bibinfo{person}{Marian Verhelst}, {and} \bibinfo{person}{Tobias Grosser}.} \bibinfo{year}{2025}\natexlab{}.
\newblock \showarticletitle{A Multi-level Compiler Backend for Accelerated Micro-kernels Targeting RISC-V ISA Extensions}. In \bibinfo{booktitle}{\emph{Proceedings of the 23rd ACM/IEEE International Symposium on Code Generation and Optimization}} (Las Vegas, NV, USA) \emph{(\bibinfo{series}{CGO '25})}. \bibinfo{publisher}{Association for Computing Machinery}, \bibinfo{address}{New York, NY, USA}, \bibinfo{pages}{163–178}.
\newblock
\urldef\tempurl%
\url{https://doi.org/10.1145/3696443.3708952}
\showDOI{\tempurl}


\bibitem[L{\"u}cke et~al\mbox{.}(2025)]%
        {lucke2025transform}
\bibfield{author}{\bibinfo{person}{Martin~Paul L{\"u}cke}, \bibinfo{person}{Oleksandr Zinenko}, \bibinfo{person}{William~S. Moses}, \bibinfo{person}{Michel Steuwer}, {and} \bibinfo{person}{Albert Cohen}.} \bibinfo{year}{2025}\natexlab{}.
\newblock \showarticletitle{The {MLIR} Transform Dialect: Your Compiler Is More Powerful Than You Think}. In \bibinfo{booktitle}{\emph{Proceedings of the 23rd ACM/IEEE International Symposium on Code Generation and Optimization (CGO)}}. \bibinfo{publisher}{ACM}.
\newblock
\urldef\tempurl%
\url{https://doi.org/10.1145/3696443.3708922}
\showDOI{\tempurl}


\bibitem[Moses et~al\mbox{.}(2021)]%
        {moses2021polygeist}
\bibfield{author}{\bibinfo{person}{William~S. Moses}, \bibinfo{person}{Lorenzo Chelini}, \bibinfo{person}{Ruizhe Zhao}, {and} \bibinfo{person}{Oleksandr Zinenko}.} \bibinfo{year}{2021}\natexlab{}.
\newblock \showarticletitle{Polygeist: Raising C to Polyhedral MLIR}. In \bibinfo{booktitle}{\emph{Proceedings of the 30th International Conference on Parallel Architectures and Compilation Techniques (PACT)}}. \bibinfo{publisher}{IEEE}, \bibinfo{pages}{45--59}.
\newblock
\urldef\tempurl%
\url{https://doi.org/10.1109/PACT52795.2021.00011}
\showDOI{\tempurl}


\bibitem[{OpenMP Architecture Review Board}(2026)]%
        {openmp_spec}
\bibfield{author}{\bibinfo{person}{{OpenMP Architecture Review Board}}.} \bibinfo{year}{2026}\natexlab{}.
\newblock \bibinfo{title}{OpenMP Application Programming Interface Version 6.0}.
\newblock
\newblock
\urldef\tempurl%
\url{https://www.openmp.org/specifications/}
\showURL{%
\tempurl}


\bibitem[Platform(2024)]%
        {pulp-llvm}
\bibfield{author}{\bibinfo{person}{PULP Platform}.} \bibinfo{year}{2024}\natexlab{}.
\newblock \bibinfo{title}{PULP LLVM Toolchain}.
\newblock \bibinfo{howpublished}{\url{https://github.com/pulp-platform/llvm-project/tree/release/18.x}}.
\newblock
\newblock
\shownote{Accessed: 2026-04-01}.


\bibitem[Platform(2026a)]%
        {pulp-gcc}
\bibfield{author}{\bibinfo{person}{PULP Platform}.} \bibinfo{year}{2026}\natexlab{a}.
\newblock \bibinfo{title}{pulp-riscv-gnu-toolchain}.
\newblock \bibinfo{howpublished}{\url{https://github.com/pulp-platform/pulp-riscv-gnu-toolchain}}.
\newblock
\newblock
\shownote{Accessed: 2026-04-01}.


\bibitem[Platform(2026b)]%
        {pulp-sdk}
\bibfield{author}{\bibinfo{person}{PULP Platform}.} \bibinfo{year}{2026}\natexlab{b}.
\newblock \bibinfo{title}{PULP SDK}.
\newblock \bibinfo{howpublished}{\url{https://github.com/pulp-platform/pulp-sdk}}.
\newblock
\newblock
\shownote{Accessed: 2026-04-01}.


\bibitem[Pouchet and Yuki(2016)]%
        {polybench}
\bibfield{author}{\bibinfo{person}{Louis-No{\"e}l Pouchet} {and} \bibinfo{person}{Tomofumi Yuki}.} \bibinfo{year}{2016}\natexlab{}.
\newblock \bibinfo{title}{PolyBench/C Benchmark Suite}.
\newblock \bibinfo{howpublished}{\url{http://web.cse.ohio-state.edu/~pouchet/software/polybench/}}.
\newblock
\newblock
\shownote{Accessed: 2026-04-01}.


\bibitem[Rossi et~al\mbox{.}(2014)]%
        {rossi2014energy}
\bibfield{author}{\bibinfo{person}{Davide Rossi}, \bibinfo{person}{Igor Loi}, \bibinfo{person}{Francesco Conti}, \bibinfo{person}{Giuseppe Tagliavini}, \bibinfo{person}{Antonio Pullini}, {and} \bibinfo{person}{Andrea Marongiu}.} \bibinfo{year}{2014}\natexlab{}.
\newblock \showarticletitle{Energy efficient parallel computing on the PULP platform with support for OpenMP}. In \bibinfo{booktitle}{\emph{2014 IEEE 28th Convention of Electrical \& Electronics Engineers in Israel (IEEEI)}}. IEEE, \bibinfo{publisher}{IEEE}, \bibinfo{address}{Eilat, Israel}, \bibinfo{pages}{1--5}.
\newblock


\bibitem[Schardl et~al\mbox{.}(2017)]%
        {tapir}
\bibfield{author}{\bibinfo{person}{Tao~B. Schardl}, \bibinfo{person}{William~S. Moses}, {and} \bibinfo{person}{Charles~E. Leiserson}.} \bibinfo{year}{2017}\natexlab{}.
\newblock \showarticletitle{Tapir: Embedding Fork-Join Parallelism into LLVM's Intermediate Representation}. In \bibinfo{booktitle}{\emph{Proceedings of the 22nd ACM SIGPLAN Symposium on Principles and Practice of Parallel Programming (PPoPP)}} (Austin, Texas, USA) \emph{(\bibinfo{series}{PPoPP '17})}. \bibinfo{publisher}{Association for Computing Machinery}, \bibinfo{address}{New York, NY, USA}, \bibinfo{pages}{249–265}.
\newblock
\urldef\tempurl%
\url{https://doi.org/10.1145/3018743.3018758}
\showDOI{\tempurl}


\bibitem[Wang et~al\mbox{.}(2023)]%
        {wang2023upir}
\bibfield{author}{\bibinfo{person}{Anjia Wang}, \bibinfo{person}{Xinyao Yi}, {and} \bibinfo{person}{Yonghong Yan}.} \bibinfo{year}{2023}\natexlab{}.
\newblock \showarticletitle{UPIR: Toward the Design of Unified Parallel Intermediate Representation for Parallel Programming Models}. In \bibinfo{booktitle}{\emph{Proceedings of the International Conference on Parallel Architectures and Compilation Techniques}} (Chicago, Illinois) \emph{(\bibinfo{series}{PACT '22})}. \bibinfo{publisher}{Association for Computing Machinery}, \bibinfo{address}{New York, NY, USA}, \bibinfo{pages}{530–531}.
\newblock
\urldef\tempurl%
\url{https://doi.org/10.1145/3559009.3569646}
\showDOI{\tempurl}


\bibitem[Wang et~al\mbox{.}(2013)]%
        {wang2013libeomp}
\bibfield{author}{\bibinfo{person}{Cheng Wang}, \bibinfo{person}{Sunita Chandrasekaran}, \bibinfo{person}{Barbara Chapman}, {and} \bibinfo{person}{Jim Holt}.} \bibinfo{year}{2013}\natexlab{}.
\newblock \showarticletitle{libEOMP: a portable OpenMP runtime library based on MCA APIs for embedded systems}. In \bibinfo{booktitle}{\emph{Proceedings of the 2013 International Workshop on Programming Models and Applications for Multicores and Manycores}} (Shenzhen, Guangdong, China) \emph{(\bibinfo{series}{PMAM '13})}. \bibinfo{publisher}{Association for Computing Machinery}, \bibinfo{address}{New York, NY, USA}, \bibinfo{pages}{83–92}.
\newblock
\urldef\tempurl%
\url{https://doi.org/10.1145/2442992.2443001}
\showDOI{\tempurl}


\end{thebibliography}

\end{document}